\documentclass[floatfix,twocolumn,showpacs,preprintnumbers,amsmath,amssymb,pra,superscriptaddress,longbibliography]{revtex4-1}
\usepackage{color}
\usepackage[usenames,dvipsnames,svgnames,table]{xcolor}
\usepackage[colorlinks=true,linkcolor=blue,urlcolor=blue,citecolor=blue]{hyperref}
\usepackage{mathtools}
\usepackage{graphicx}
\usepackage{dcolumn}
\usepackage{array}
\usepackage{lipsum}
\usepackage{bm}
\usepackage{subfigure}
\usepackage{amssymb}
\usepackage{multirow}
\usepackage{tabularx}
\usepackage{amsmath}
\usepackage{braket}
\usepackage{csquotes}
\graphicspath{{plots/}}
 \usepackage{lipsum}
\usepackage{mathrsfs}
\usepackage{MnSymbol}
	
\newcommand{\beq}{\begin{equation}}
\newcommand{\eeq}{\end{equation}}
\newcommand{\bea}{\begin{eqnarray}}
\newcommand{\eea}{\end{eqnarray}}

\begin{document}
\title{Application of a spherically averaged pair potential in \emph{ab initio} path integral Monte Carlo simulations of the warm dense electron gas}
\author{Tobias Dornheim}
\email{t.dornheim@hzdr.de}
\affiliation{Center for Advanced Systems Understanding (CASUS), D-02826 G\"orlitz, Germany}
\affiliation{Helmholtz-Zentrum Dresden-Rossendorf (HZDR), D-01328 Dresden, Germany}
\author{Thomas~M.~Chuna}
\affiliation{Center for Advanced Systems Understanding (CASUS), D-02826 G\"orlitz, Germany}
\affiliation{Helmholtz-Zentrum Dresden-Rossendorf (HZDR), D-01328 Dresden, Germany}
\author{Hannah M.~Bellenbaum}
\affiliation{Center for Advanced Systems Understanding (CASUS), D-02826 G\"orlitz, Germany}
\affiliation{Helmholtz-Zentrum Dresden-Rossendorf (HZDR), D-01328 Dresden, Germany}
\affiliation{Institut f\"ur Physik, Universit\"at Rostock, D-18057 Rostock, Germany}
\author{Zhandos~A.~Moldabekov}
\affiliation{Center for Advanced Systems Understanding (CASUS), D-02826 G\"orlitz, Germany}
\affiliation{Helmholtz-Zentrum Dresden-Rossendorf (HZDR), D-01328 Dresden, Germany}
\author{Panagiotis Tolias}
\affiliation{Space and Plasma Physics, Royal Institute of Technology (KTH), Stockholm, SE-100 44, Sweden}
\author{Jan Vorberger}
\affiliation{Helmholtz-Zentrum Dresden-Rossendorf (HZDR), D-01328 Dresden, Germany}
\begin{abstract}
Spherically averaged periodic pair potentials offer the enticing promise to provide accurate results at a drastically reduced computational cost compared to the traditional Ewald sum. In this work, we employ the pair potential by Yakub and Ronchi [\textit{J.~Chem.~Phys.}~\textbf{119}, 11556 (2003)] in \emph{ab initio} path integral Monte Carlo (PIMC) simulations of the warm dense uniform electron gas. Overall, we find very accurate results with respect to Ewald reference data for integrated properties such as the kinetic and potential energy, whereas wavenumber resolved properties such as the static structure factor $S(\mathbf{q})$, the static linear density response $\chi(\mathbf{q})$ and the static quadratic density response $\chi^{(2)}(\mathbf{q},0)$ fluctuate for small $q$. In addition, we perform an analytic continuation to compute the dynamic structure factor $S(\mathbf{q},\omega)$ from PIMC results of the imaginary-time density--density correlation function $F(\mathbf{q},\tau)$ for both pair potentials. Our results have important implications for future PIMC calculations, which can be sped up significantly using the YR potential for the estimation of equation-of-state properties or $q$-resolved observables in the non-collective regime, whereas a full Ewald treatment is mandatory to accurately resolve physical effects manifesting for smaller $q$, including the evaluation of compressibility sum rules, the interpretation of x-ray scattering experiments at small scattering angles, and the estimation of optical and transport properties.
\end{abstract}
\maketitle

\section{Introduction}

Warm dense matter (WDM) comprises a complex state of matter that is often characterized by at least two dimensionless parameters simultaneously of the order of one~\cite{wdm_book,Ott2018}: the density parameter---also coined as Wigner-Seitz radius---$r_s=(3/4\pi n)^{1/3}$, with $n$ being the electronic number density, and the degeneracy temperature $\Theta=(\beta E_\textnormal{F})^{-1}$, where $\beta=1/k_\textnormal{B}T$ is the inverse temperature in energy units and $E_\textnormal{F}$ the usual Fermi energy of the electrons~\cite{quantum_theory}.
In nature, such seemingly extreme conditions abound in astrophysical objects such as giant planet interiors~\cite{Benuzzi_Mounaix_2014,wdm_book,drake2018high} and white dwarf atmospheres~\cite{Kritcher2020,SAUMON20221}.
In addition, extreme states of matter become increasingly important for cutting-edge technological applications, with inertial confinement fusion (ICF)~\cite{Betti2016,Hurricane_RevModPhys_2023,drake2018high} being a prime example. Indeed, recent spectacular achievements at the National Ignition Facility (NIF)~\cite{AbuShawareb_PRL_2024,Zylstra2022} and the OMEGA laser facility~\cite{Gopalaswamy2024} have demonstrated the principal feasibility of using ICF to produce net energy, but the pathway towards a commercial utilization of this technology will require a further optimization of current energy gains by several orders of magnitude~\cite{Batani_roadmap}. This, in turn, will require integrated simulations with real predictive capability, which must also cover the initial stages of the compression path, where both the fusion fuel and its surrounding ablator material have to traverse the WDM regime in a controlled way~\cite{hu_ICF}.

From a theoretical perspective, the accurate description of WDM poses a formidable challenge as it must treat holistically a plethora of physical effects. Specifically, the moderate to strong coupling of the Coulomb interacting electrons in nuclei often rules out weak-coupling expansions such as equilibrium or nonequilibrium Greens functions~\cite{bonitz2024principles,Schoof_PRL_2015,kwong_prl-00}. In addition, quantum degeneracy and quantum diffraction effects prevent the application of semi-classical methods such as molecular dynamics simulations with effective quantum pair potentials between the electrons except at very high temperatures~\cite{bonitz2024principles,Golubnichy_CPP_2002}. On the other hand, the strong thermal excitations that occur when $\Theta\sim1$ generally prevent a straightforward application of the well-stocked arsenal of ground-state simulation tools, e.g.~from quantum chemistry and material science~\cite{karasiev_importance,kushal,Sjostrom_PRB_2014}. Finally, WDM states often feature partial ionization rendering them substantially more complex than either the well defined orbitals of bound electrons at ambient conditions or the fully ionized semi-classical plasma.

In this situation, \emph{ab initio} thermal density functional theory (DFT)~\cite{Mermin_DFT_1965} has emerged as the de-facto workhorse of WDM theory as it often balances a reasonable level of accuracy with an acceptable computational effort. Specifically, the combination of a DFT treatment of the electrons with a classical or quantum molecular dynamics propagation of the ions within the widely assumed Born-Oppenheimer approximation~\cite{wdm_book} allows for the estimation of a broad range of material properties such as the equation-of-state, ionic correlation functions, and electronic transport properties; see the review article by Bonitz \textit{et al.}~\cite{bonitz2024principles} for a topical overview. The accuracy of a given DFT simulation decisively relies on the employed electronic exchange--correlation (XC) functional, which has to be supplied as an external input. At ambient conditions, where electrons can generally be assumed to be in their respective ground state, the applicability of the zoo of existing functionals~\cite{Goerigk_PCCP_2017} for different materials and properties is relatively well understood. However, it is clear that thermal DFT simulations of real WDM systems require a thermal XC functional that explicitly depends on the temperature. Here, the situations is significantly less clear than in the extensively explored ground state limit, as available XC-functionals are substantially less numerous~\cite{ksdt,Sjostrom_PRB_2014,groth_prl,Karasiev_PRL_2018,Karasiev_PRB_2022}, and their respective accuracy has been benchmarked in far fewer situations~\cite{Moldabekov_JCTC_2024, PPNP_2025, Moldabekov_JPCL_2023, Moldabekov_JCP_2023, Moldabekov_JCTC_2023}.

In practice, both the construction of thermal XC functionals by itself as well as their rigorous benchmarking require the availability of highly accurate results that have been obtained without the need for any empirical external input. In principle, experimental observations would constitute a possible route, but their respective interpretation is notoriously difficult~\cite{Kasim_POP_2019} and often depends on a number of uncontrolled model assumptions such as the decomposition of the electrons into effectively bound and free states~\cite{Gregori_PRE_2003,dornheim2024modelfreerayleighweightxray,Dornheim_T_2022,Tilo_Nature_2023,boehme2023evidence,dharmawardana2025xraythomsonscatteringstudies}.
A more realistic option is given by first-principles quantum Monte Carlo (QMC) methods~\cite{anderson2007quantum}, with the \emph{ab initio} path integral Monte Carlo (PIMC) technique~\cite{cep} being the method of choice at finite temperatures. Being exact within the given Monte Carlo error bars, the central limitation of PIMC simulations is the infamous fermion sign problem~\cite{troyer,Loh_PRB_1990,dornheim_sign_problem} that emerges from the antisymmetry of the fermionic density matrix under the exchange of particle coordinates. The sign problem manifests as an exponentially decreasing signal-to-noise ratio in PIMC simulations when the system size $N$ or inverse temperature $\beta$ are being increased.

Due to the pressing need to understand the properties of WDM---as well as a host of other interesting thermal Fermi-Dirac systems such as ultracold $^3$He~\cite{Ceperley_PRL_1992,Dornheim_SciRep_2022,morresi2024normalliquid3hestudied} or electrons in quantum dots~\cite{PhysRevLett.82.3320,Reimann_RevModPhys_2002,Dornheim_NJP_2022}---there has been a remarkable surge of progress in fermionic QMC simulations over the last 15 years or so~\cite{Driver_PRL_2012,Brown_PRL_2013,Blunt_PRB_2014,Schoof_PRL_2015,Malone_JCP_2015,Dornheim_NJP_2015,Groth_PRB_2016,Malone_PRL_2016,Yilmaz_JCP_2020,Hirshberg_JCP_2020,Dornheim_Bogoliubov_2020,Joonho_JCP_2021,Hou_PRB_2022,Xiong_JCP_2022,Dornheim_JCP_xi_2023,dornheim2024directfreeenergycalculation}. In fact, recent methodological developments based on path integral calculations of fictitious identical particles~\cite{Xiong_JCP_2022,Dornheim_JCP_xi_2023,xiong2024gpuaccelerationabinitio,dornheim2025fermionicfreeenergiestextitab} allow for the simulation of up to $N=1000$ electrons at weak to moderate levels of quantum degeneracy~\cite{Dornheim_JPCL_2024}.

These developments open up both promising opportunities and new challenges. On the one hand, large simulations reduce finite-size effects, i.e., the difference in simulation results between a finite $N$ and the thermodynamic limit (TDL) of $N\to\infty$ at a constant density parameter $r_s$~\cite{Fraser_PRB_1996,Levashov_CPP_2024,dornheim2024chemicalpotentialwarmdense}. This is particularly important for multi-component systems consisting of both electrons and nuclei, for which finite-size corrections are substantially less developed compared to the more simple uniform electron gas (UEG)~\cite{review,Chiesa_PRL_2006,Drummond_PRB_2008,dornheim_prl,Dornheim_JCP_2021,Holzmann_PRB_2016}. In addition, a large box length $L$ (assuming a cubical simulation cell of volume $\Omega=L^3$) allows to access small wavenumbers $q$, since the minimum wavenumber is given by $q_\textnormal{min}=2\pi/L$ due to the momentum discretization. This is crucial for the description of x-ray scattering experiments in a forward scattering geometry~\cite{Glenzer_PRL_2007,Gawne_PRB_2024,Sperling_PRL_2015,Bellenbaum_APL_2025,Gawne_PRB_2024} that probe collective modes such as the plasmon oscillation~\cite{Hamann_CPP_2020}, and might eventually even facilitate the estimation of optical properties that are defined in the limit of $q\to0$~\cite{bonitz2024principles}. On the other hand, PIMC simulations of quantum many-body systems with $N=1000$ (and $P\sim10^2$ replicas of each particle on different imaginary-time slices, see below) and beyond become computationally demanding even without the exponential bottleneck due to the sign problem. Specifically, the main computational expense is given by the evaluation of the long-range pair interaction, which is typically treated with the usual Ewald sum over all periodic images~\cite{Fraser_PRB_1996}.

A vastly cheaper alternative is given by spherically averaged pair potentials, that have been averaged over different angular orientations of the periodic array of images of the particles in the main simulation cell~\cite{Yakub_JCP_2003,Yakub2005,Vernizzi_PRE_2011,Fukuda2012,Demyanov_2022,Levashov_CPP_2024}. Here, we explore the spherically averaged pair potential introduced by Yakub and Ronchi (YR)~\cite{Yakub_JCP_2003,Yakub2005}. Demyanov \textit{et al.}~\cite{Levashov_CPP_2024} applied the YR potential to Monte Carlo simulations of the classical one-component plasma that consists of negative unit charges in a homogeneous neutralizing background and found a comparable convergence of the internal energy to the standard Ewald potential. In addition, V.~Filinov \emph{et al.}~\cite{Filinov_PRE_2015,Filinov_PRE_2020,Filinov_JPA_2022} used the YR potential to compute the internal energy (and its correlation contribution) in PIMC simulations of the UEG; the quantum analogue of the classical one-component plasma. Very recently, A.~Filinov and Bonitz~\cite{Filinov_PRE_2023} tested the Ewald and the YR potentials in PIMC simulations of warm dense hydrogen and ``... \emph{found that, for the simulation parameters used in this paper, the results for both are practically indistinguishable}''. However, these references were limited to integrated properties such as energies and pressure, sometimes extending to spatial observables such as cluster analyses. The effect of the pair potential onto wavenumber resolved properties has, thus, remained unclear. This is unfortunate for (at least) two reasons: (i) the invariance of the electronic static structure factor $S(\mathbf{q})$ with respect to the system size even for $N\sim10$ is of key importance for finite-size corrections e.g.~of the interaction energy $W$~\cite{dornheim_prl,Chiesa_PRL_2006} and has been pivotal for the construction of accurate parametrizations of the UEG in the TDL~\cite{dornheim_prl,review,ksdt,Karasiev_PRL_2018}; (ii) the invariance with respect to $N$ of species-resolved static structure factors $S_{ab}(\mathbf{q})$ and related imaginary-time correlation functions (ITCF) $F_{ab}(\mathbf{q},\tau)$, where $t=-i\hbar\tau$ is the imaginary time with $\tau\in[0,\beta]$, in PIMC simulations of real WDM systems such as hydrogen~\cite{Dornheim_JCP_2024,Dornheim_MRE_2024} and beryllium~\cite{Dornheim_Science_2024,dornheim2024modelfreerayleighweightxray} facilitates the comparison with experimental observations, most notably x-ray Thomson scattering (XRTS) measurements~\cite{Tilo_Nature_2023,Dornheim_Science_2024}.

In the present work, we remedy this situation by presenting a rigorous and detailed analysis of the effect of the YR potential compared to the standard Ewald sum onto different properties of the UEG, including different energies, static structure factor, and linear / quadratic static density response functions~\cite{dornheim_ML,Dornheim_JCP_ITCF_2021}. Finally, we carry out an analytic continuation~\cite{dornheim_dynamic,chuna2025dualformulationmaximumentropy,JARRELL1996133} to compute the dynamic structure factor $S(\mathbf{q},\omega)$ for selected wavenumbers. As a general trend, we find that the standard Ewald and YR potentials do indeed give comparable results for integrated properties such as the kinetic and potential energy. However, use of the YR potential breaks the $N$-invariance of $q$-resolved properties (in particular at small wavenumbers), which constitutes a serious limitation for comparisons with experiments and finite-size corrections.

The paper is organized as follows: In Sec.~\ref{sec:theory}, we provide the theoretical background with a focus on the Ewald and YR pair potentials. In Sec.~\ref{sec:results}, we present our new PIMC simulation results on the warm dense UEG at $r_s=3.23$ (Sec.~\ref{sec:WDM}), the electron liquid at $r_s=10$ (Sec.~\ref{sec:electron_liquid}),  the analytic continuation to obtain the dynamic structure factor (Sec.~\ref{sec:anal_cont}), and the static quadratic density response $\chi^{(2)}(\mathbf{q},0)$ (Sec.~\ref{sec:quadratic}). The paper is concluded by a summary and outlook in Sec.~\ref{sec:outlook}.

\begin{figure}\centering
\includegraphics[width=0.45\textwidth]{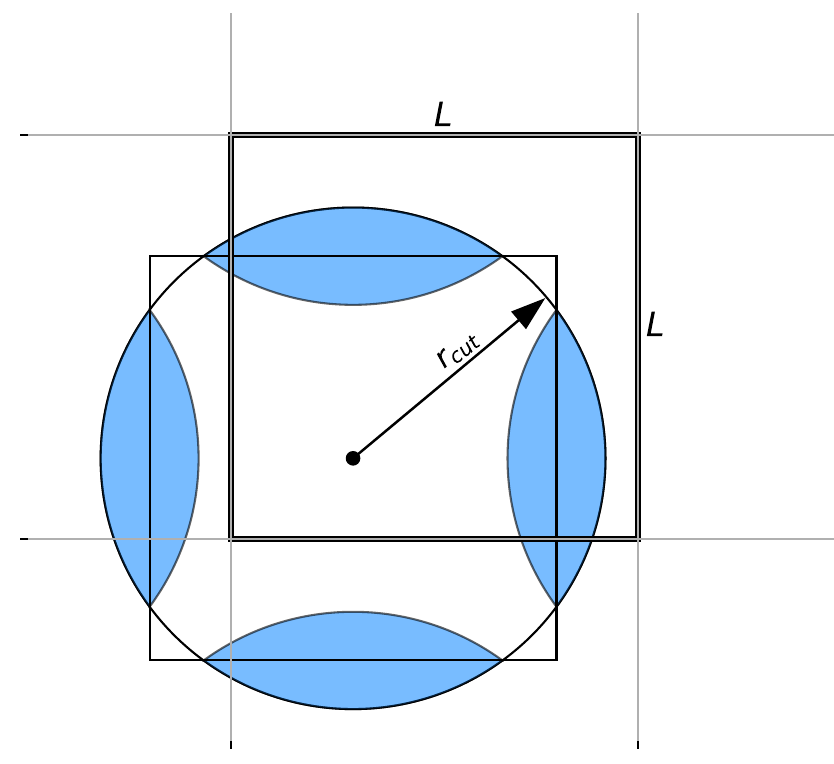}
\caption{\label{fig:scheme} 
Schematic illustration of the YR potential, Eq.~(\ref{eq:Yakub_potential}). A random particle in the main simulation cell of length $L$ interacts with all particles and their periodic images within the volume equivalent sphere of radius $r_\textnormal{cut}$. This leads to a mandated double counting of both an original particle and its image in the shaded blue areas. Adapted from Ref.~\cite{Yakub_JCP_2003}.
}
\end{figure}

\section{Theory\label{sec:theory}}

The \emph{ab initio} PIMC method~\cite{cep} is based on the celebrated quantum-to-classical mapping~\cite{Chandler_JCP_1981} and has been described extensively in the literature~\cite{cep,boninsegni2,Dornheim_permutation_cycles}. We limit ourselves to a brief summary of our implementation and its features. We assume Hartree atomic units.

The main advantage of PIMC is its capability to provide quasi-exact (i.e., exact within the given Monte Carlo error bars) results for any thermodynamic observable for the specified quantum $N$-body problem. In practice, an important convergence parameter is given by the number of imaginary-time propagators $P$. Here, we use the primitive high-temperature factorization $e^{-\epsilon\hat{H}}\approx e^{-\epsilon\hat{K}}e^{-\epsilon\hat{V}}$, with $\epsilon=\beta/P$ and $\hat{K}$, $\hat{V}$ the kinetic and potential contributions to the full Hamiltonian $\hat{H}$. The convergence with $P$ is discussed in Sec.~\ref{sec:results}. To ensure ergodic sampling of all relevant path configurations, we use the canonical extended ensemble adaption of the worm algorithm by Boninsegni \emph{et al.}~\cite{boninsegni1,boninsegni2} presented in Ref.~\cite{Dornheim_PRB_nk_2021}, which is implemented in the open-source \texttt{ISHTAR} code~\cite{ISHTAR}.
Finally, we consider electrons, which obey Fermi-Dirac statistics. The anti-symmetry of the thermal density matrix under the exchange of particle coordinates is taken into account by sampling the permutation structure~\cite{Dornheim_permutation_cycles} without any nodal restrictions. Hence, our simulations are computationally costly due to the fermion sign problem~\cite{dornheim_sign_problem}, but exact within error bars. Moreover, we retain full access to equilibrium system dynamics in the form of various imaginary-time correlation functions~\cite{Dornheim_JCP_ITCF_2021}. 

We consider $N=N^\uparrow+N^\downarrow$ (with $N^\uparrow=N^\downarrow$) electrons in a cubic simulation cell of volume $\Omega=L^3$ and standard periodic boundary conditions. To take into account the long-range nature of the Coloumb interaction, we consider the usual infinite array of periodic images; the corresponding solution to Poisson's equation is then given by the Ewald sum~\cite{Fraser_PRB_1996}.
Following the notation by Fraser \emph{et al.}~\cite{Fraser_PRB_1996}, we express the Ewald pair interaction as
\begin{widetext}
\begin{eqnarray}\label{eq:Ewald_potential}
    \phi_\textnormal{E}(\mathbf{r}_a,\mathbf{r}_b) = \frac{1}{\Omega} \sum_{\mathbf{G}\neq\mathbf{0}} \frac{\textnormal{exp}\left(i2\pi\mathbf{G}\cdot(\mathbf{r}_a-\mathbf{r}_b) - \frac{\pi^2\mathbf{G^2}}{\kappa^2} \right)}{\pi \mathbf{G}^2} - \frac{\pi}{\kappa^2 \Omega} + 
    \sum_\mathbf{n} \frac{\textnormal{erfc}\left(\kappa|\mathbf{r}_a-\mathbf{r}_b+\mathbf{n}L|\right)}{|\mathbf{r}_a-\mathbf{r}_b + \mathbf{n}L|}\ ,
\end{eqnarray}
\end{widetext}
with $\mathbf{n}=(n_x,n_y,n_z)^T$ being an integer vector and where the reciprocal lattice vectors $\mathbf{G}$ have been defined \emph{without} including the customary factor of $2\pi$. We note that $\kappa$---often coined as the Ewald parameter---can be chosen freely to shift weights between the sums over $\mathbf{G}$ and $\mathbf{n}$, while the fully converged result is invariant. In practice, both sums have to be truncated, which affects the computation cost. For completeness, we note that a vast literature exists on efficient numerical implementation of Eq.~(\ref{eq:Ewald_potential}) for Monte Carlo simulations, see, e.g., Refs.~\cite{TOUKMAJI199673,RAJAGOPAL1994399,KYLANPAA201664,Janke_PRX_2023} and references therein. The full Hamiltonian of the UEG with Ewald interaction is given by 
\begin{eqnarray}\label{eq:Hamiltonian_Ewald}
    \hat{H}_\textnormal{E} = -\frac{1}{2}\sum_{l=1}^N \nabla_l^2 + \sum_{l<k}^N \phi_\textnormal{E}(\hat{\mathbf{r}}_l,\hat{\mathbf{r}}_k) + \frac{N\xi_\textnormal{M}}{2}\ ,
\end{eqnarray}
with $\xi_\textnormal{M}=-2.837297(3/4\pi)^{1/3}N^{-1/3}r_s^{-1}$ the Madelung constant~\cite{Schoof_PRL_2015} that takes into account the interaction between a charge and its own background and periodic array of self-images.

Angular averaging over all possible orientations of the infinite periodic array of images leads to a considerable simplification of Eq.~(\ref{eq:Ewald_potential}). For the one-component plasma, the Hamiltonian is given by~\cite{Levashov_CPP_2024}
\begin{eqnarray}\label{eq:Hamiltonian_Yakub}
    \hat{H}_\textnormal{YR} = -\frac{1}{2}\sum_{l=1}^N \nabla_l^2 + \sum_{l<k}^N \sum_\mathbf{n} \phi_\textnormal{YR}(\hat{\mathbf{r}}_l,\hat{\mathbf{r}}_k+\mathbf{n}L) - \frac{3}{4 r_\textnormal{cut}}\left(1+\frac{N}{5}\right)\ ,\quad
\end{eqnarray}
with the definition of the YR-potential~\cite{Yakub2005,Yakub_JCP_2003}
\begin{widetext}
\begin{eqnarray}\label{eq:Yakub_potential}
    \phi_\textnormal{YR}(\mathbf{r}_a,\mathbf{r}_b) = \begin{cases}
    \frac{1}{r_{ab}}\left\{
    1 + \frac{1}{2}\left(\frac{r_{ab}}{r_\textnormal{cut}}\right)
    \left[ \left(\frac{r_{ab}}{r_\textnormal{cut}}\right)^2 - 3 \right]
    \right\},& \text{if } r_{ab} < r_\textnormal{cut}\\
    0,              & \text{otherwise}\ ,
\end{cases}
\end{eqnarray}
\end{widetext}
where $r_{ab}=|\mathbf{r}_b-\mathbf{r}_a|$.
The cutoff radius follows from the volume-equivalent sphere of the main cell, $r_\textnormal{cut}=L(3/4\pi)^{1/3}$. We note that, by definition, the sum over periodic images $\mathbf{n}$ in Eq.~(\ref{eq:Hamiltonian_Yakub}) can be truncated for $|\mathbf{n}|\geq 1.62$; still, it is possible that the YR pair potential Eq.~(\ref{eq:Yakub_potential}) does give contributions for more than one image (cf.~the blue areas in Fig.~\ref{fig:scheme}), see also the discussion in Ref.~\cite{Yakub_JCP_2003}.

Demyanov \emph{et al.}~\cite{Levashov_CPP_2024} report an acceleration by two orders of magnitude of the YR over the Ewald potential in their classical MC simulations; we find a similar acceleration of $\mathcal{O}\left(10^{1}-10^{2}\right)$ for PIMC. We further point out that typical PIMC simulations deal with $N\sim10-100$ particles, with replicas on $P\sim100$ imaginary-time slices. This often limits the utility of sophisticated multi-pole expansions of the Ewald potential, which, despite having a favorable scaling with $N$, often start becoming more efficient for $N\sim1000$.
At the same time, the numerical evaluation of Eq.~(\ref{eq:Ewald_potential}) is usually responsible for the main fraction of the compute time in PIMC simulations of periodic Coulomb systems such as the UEG, which makes any possible acceleration highly desirable.

\section{Results\label{sec:results}}

All PIMC results presented in this work have been obtained using the open-source \texttt{ISHTAR} code~\cite{ISHTAR}, and are freely available in an online repository~\cite{repo}.

\subsection{Warm dense matter\label{sec:WDM}}

\begin{figure}\centering
\includegraphics[width=0.45\textwidth]{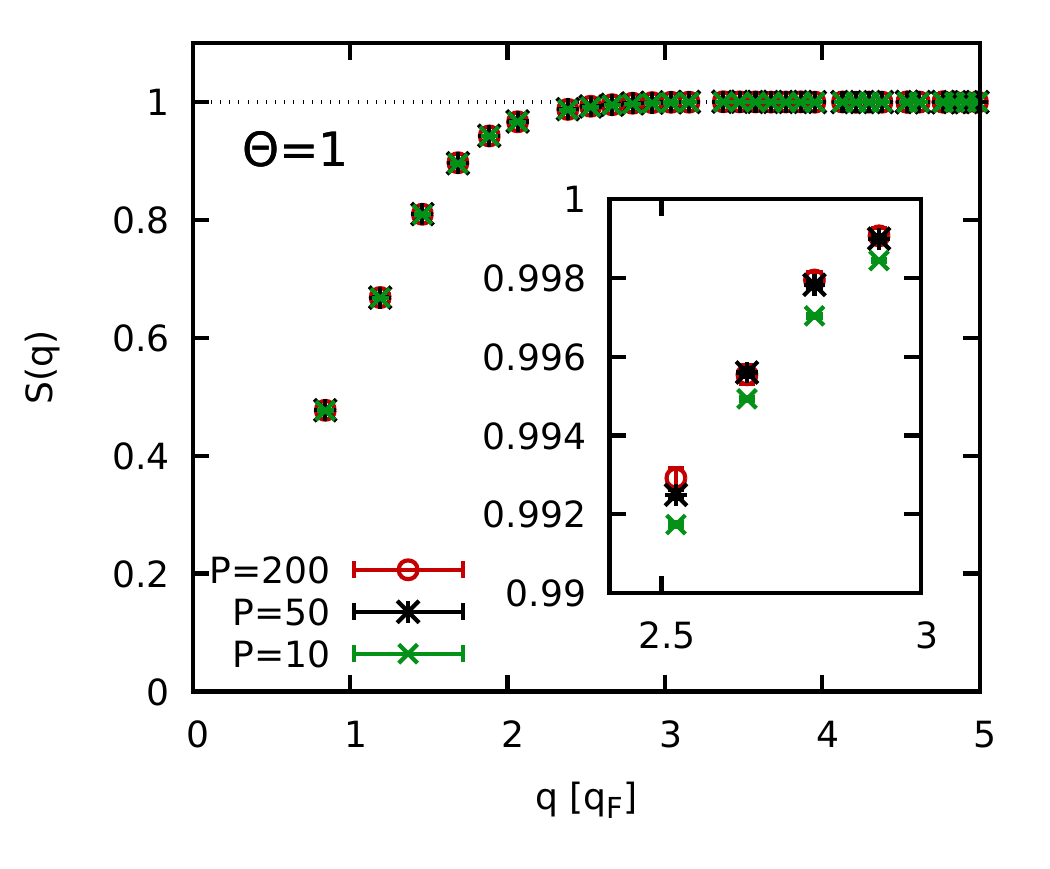}\\
\hspace*{-0.02\textwidth}\includegraphics[width=0.47\textwidth]{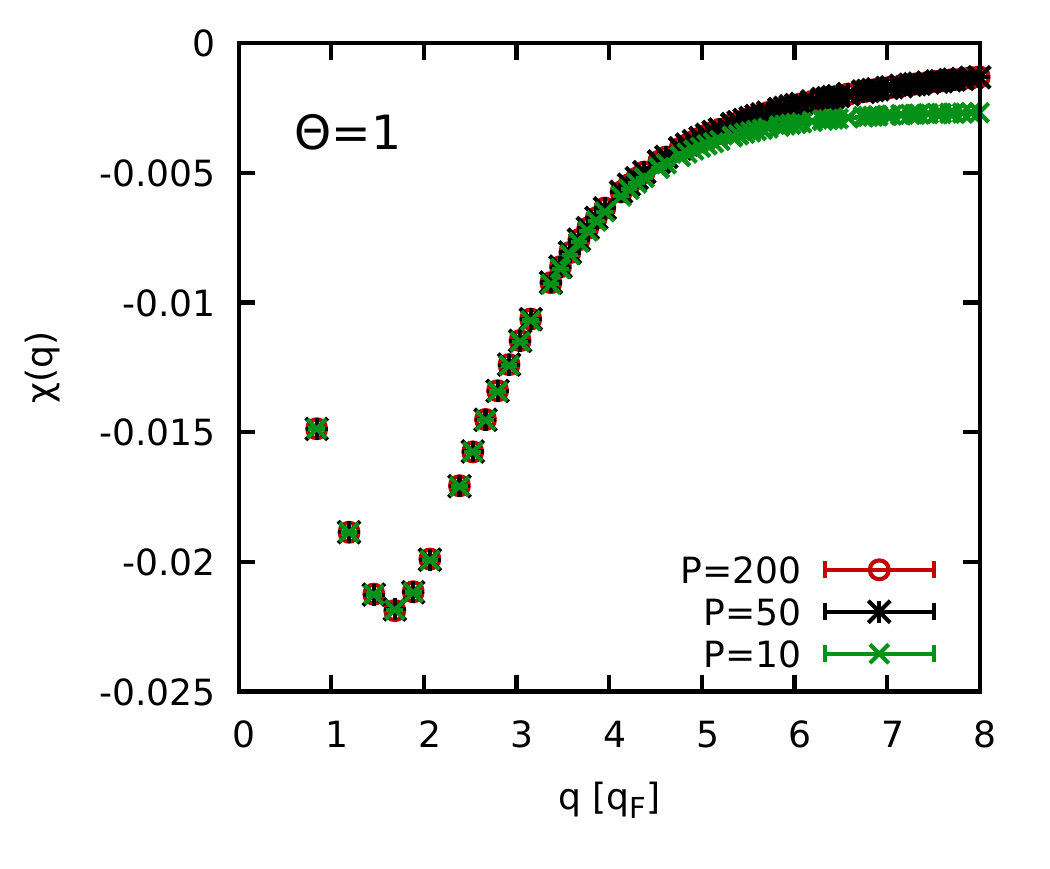}
\caption{\label{fig:P_convergence_q} PIMC results for the static structure factor $S(\mathbf{q})$ [top] and the static linear density response function $\chi(\mathbf{q})$ [bottom, cf.~Eq.~(\ref{eq:static_chi})] of the unpolarized UEG at $r_s=3.23$, $\Theta=1$, with $N=14$ and the YR pair potential. Different symbols and colors distinguish different numbers of imaginary-time propagators $P$. The top panel inset shows a magnified segment.
}
\end{figure} \begin{figure}\centering
\includegraphics[width=0.45\textwidth]{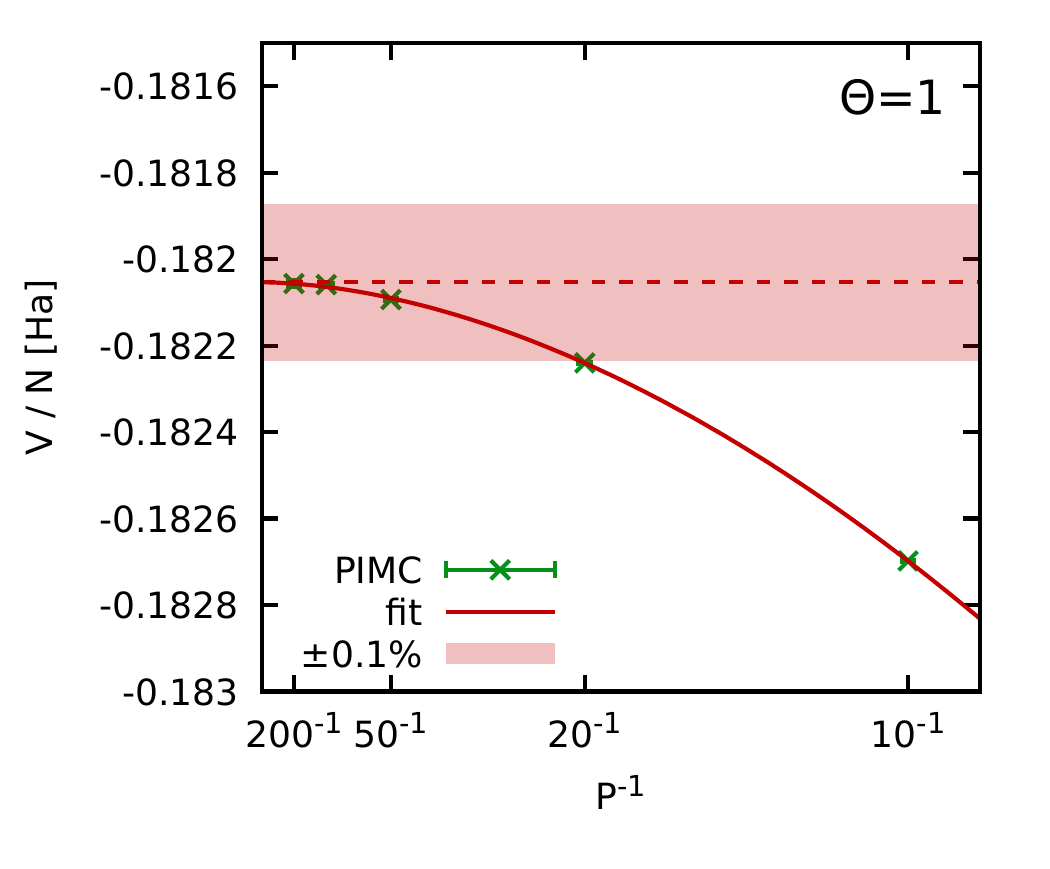}\\\includegraphics[width=0.45\textwidth]{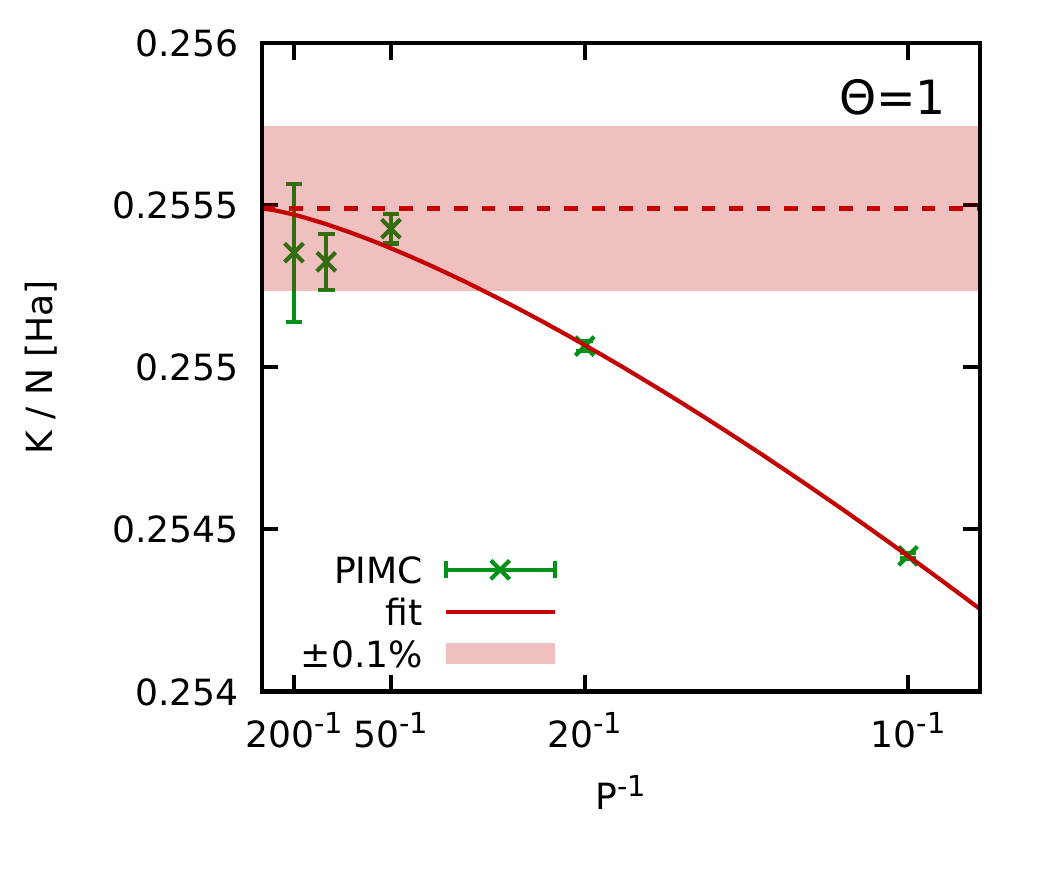}
\caption{\label{fig:P_convergence} The convergence of the potential energy $V$ (top) and kinetic energy $K$ (bottom), for $N=14$ unpolarized electrons interacting with the YR potential at $r_s=3.23$, $\Theta=1$, with the number of imaginary-time propagators $P$. The solid red lines show empirical fits of the form 
$f(x)=a + bx^c$ that have been included as a guide to the eye.}
\end{figure} 

\begin{figure*}\centering
\includegraphics[width=0.45\textwidth]{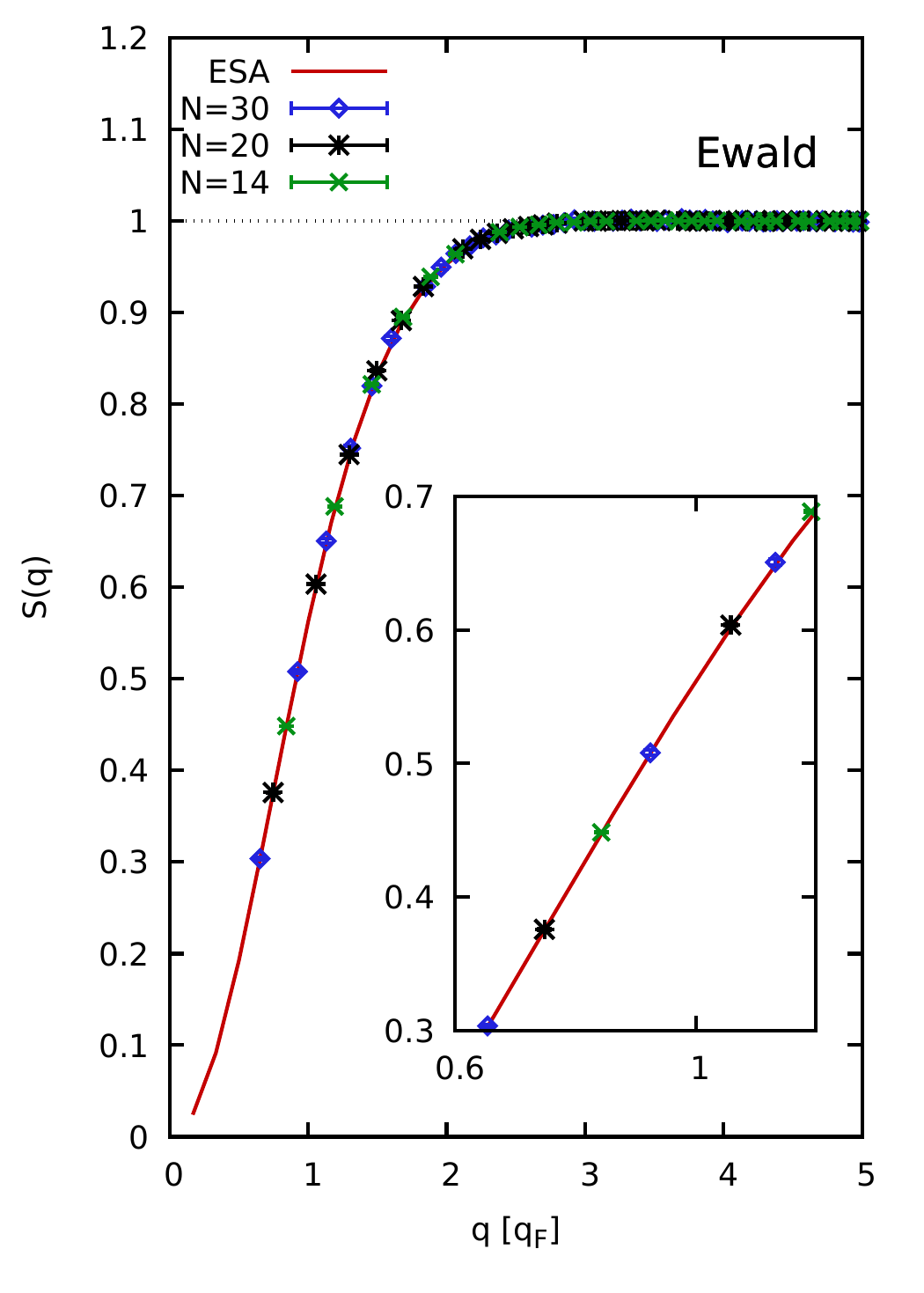}\includegraphics[width=0.45\textwidth]{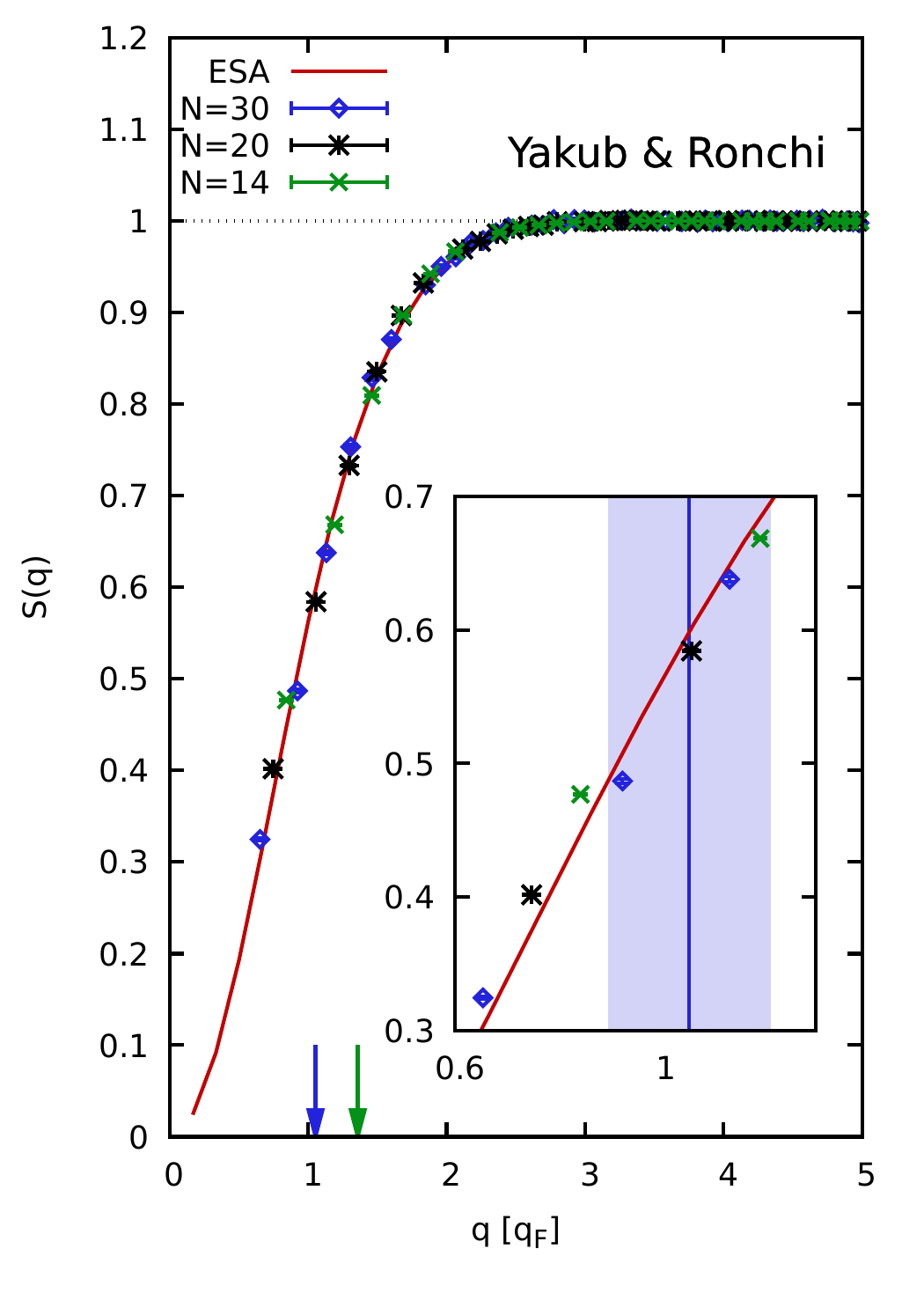}\\\includegraphics[width=0.45\textwidth]{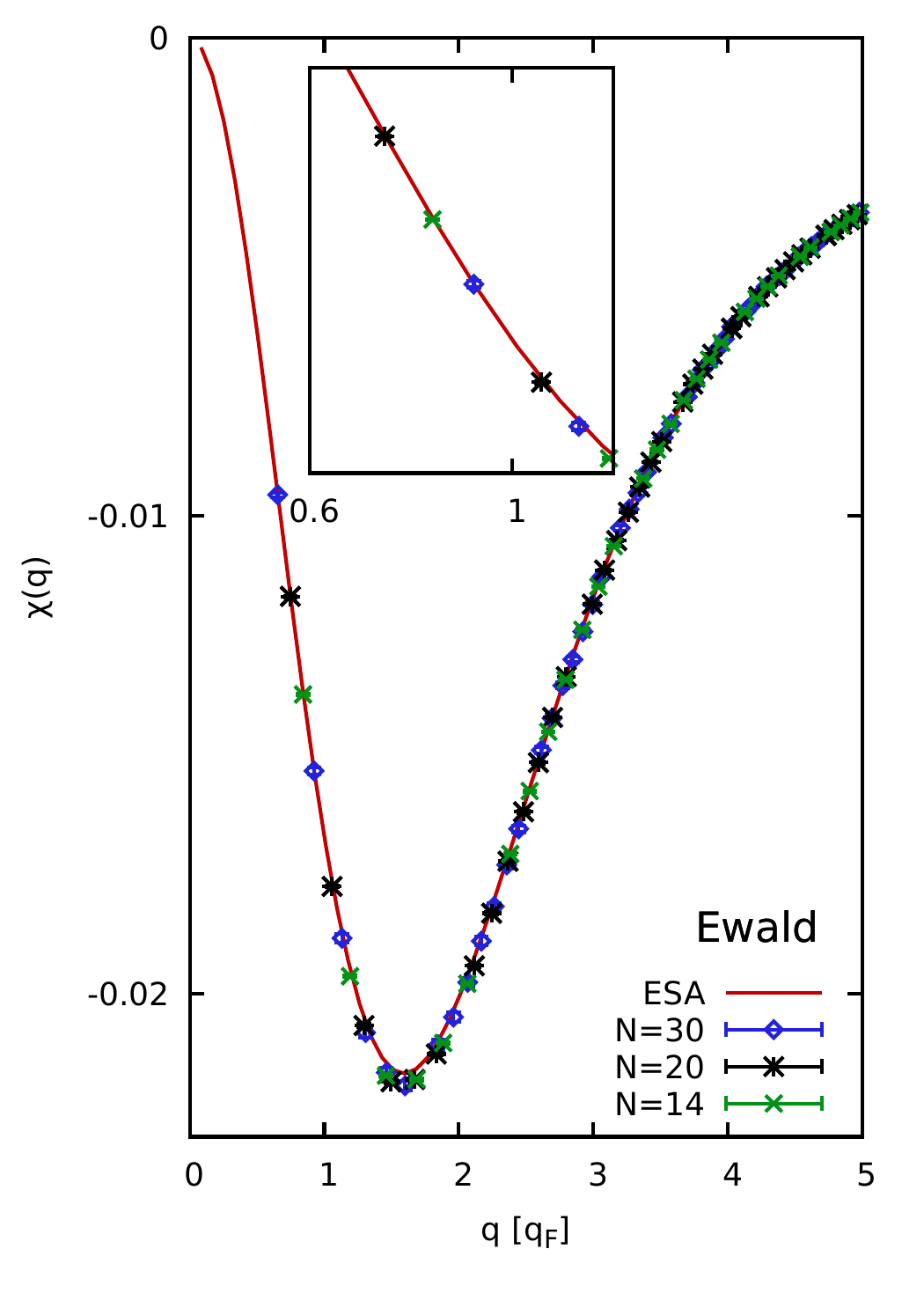}\includegraphics[width=0.45\textwidth]{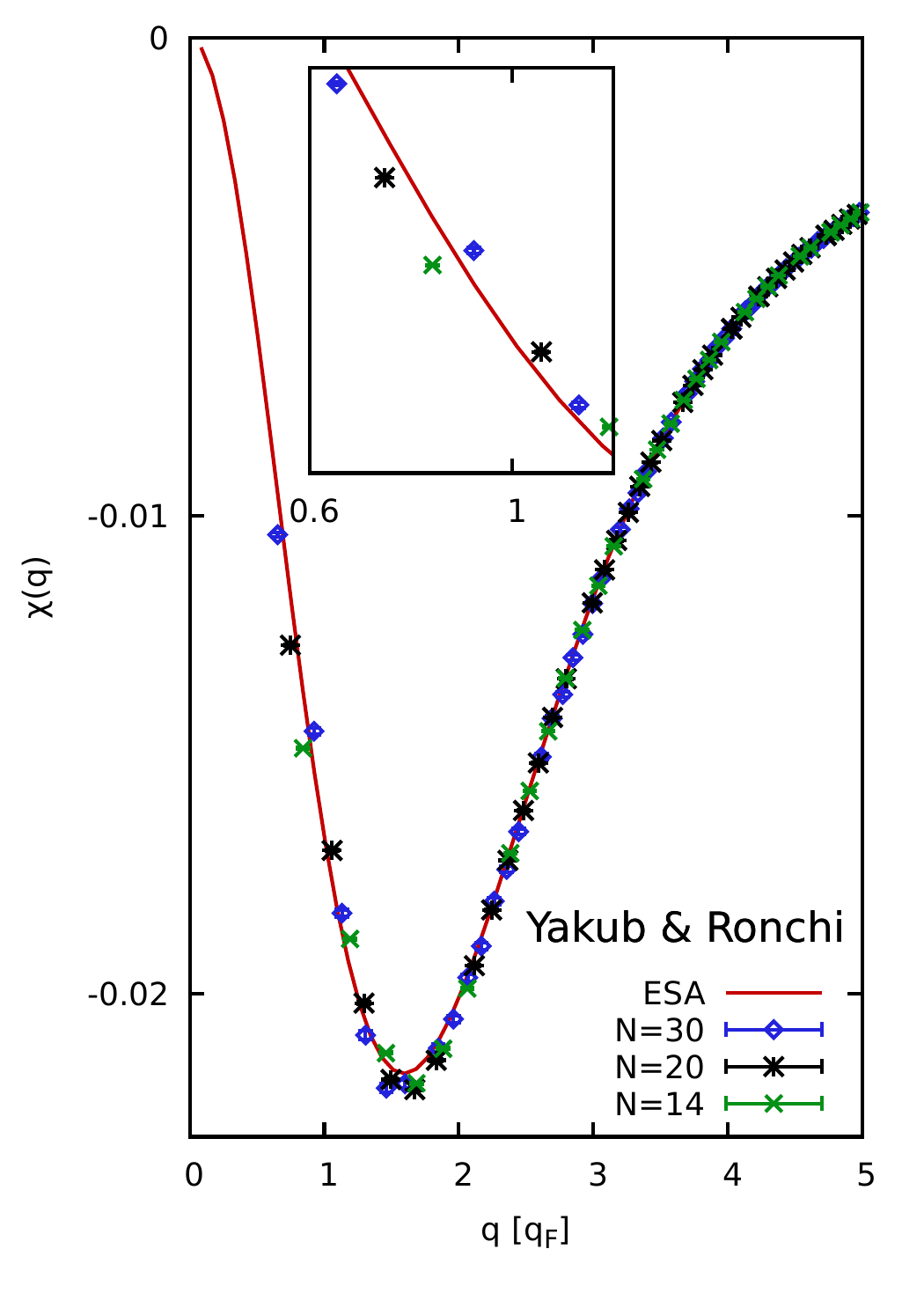}
\caption{\label{fig:SSF_rs3p23_theta1} PIMC results for the static structure factor $S(\mathbf{q})$ [top row] and the static linear density response function $\chi(\mathbf{q})$ [cf.~Eq.~(\ref{eq:static_chi}), bottom row] at $r_s=3.23$ and $\Theta=1$. Symbols: PIMC data for different $N$ computed with the Ewald potential (left) and the YR potential (right). Solid red: effective static approximation (ESA)~\cite{Dornheim_PRL_2020_ESA,Dornheim_PRB_ESA_2021}, included as a reference. The insets show magnified segments.
}
\end{figure*} 

As the first set of relevant conditions, we consider the UEG at $r_s=3.23$ and $\Theta=1$ (i.e., $T=4.8\,$eV). These fall into the WDM regime and can be realized e.g.~in experiments with optically pumped hydrogen jets~\cite{Zastrau,Fletcher_Frontiers_2022}. We note that Hamann \emph{et al.}~\cite{Hamann_PRR_2023} have predicted the emergence of a roton-type feature in the electronic dynamic structure factor of hydrogen at these parameters. This is a consequence of the partial ionization of hydrogen in this regime~\cite{bellenbaum2025estimatingionizationstatescontinuum} leading to an effectively lower free electron density, whereas the roton-type feature is absent in the UEG at $r_s=3.23$.


In the top panel of Fig.~\ref{fig:P_convergence_q}, we study the convergence of the static structure factor $S(\mathbf{q})$ (computed via the YR potential) with the number of imaginary-time propagators $P$. We only find systematic bias for $P=10$ (green crosses), whereas, e.g.~$P=50$ (black stars) and $P=200$ (red circles) cannot be distinguished within the given error bars. This can be discerned particularly well in the inset that shows a magnified segment. Indeed, even for $P=10$ the convergence error is of the order of $\sim0.1\%$.
In the bottom panel of Fig.~\ref{fig:P_convergence_q}, we repeat this analysis for the static linear density response function $\chi(\mathbf{q})$. In principle, the latter can be obtained by applying a small harmonic perturbation to the system and measuring its density response~\cite{moroni,moroni2,bowen2,dornheim_pre,groth_jcp,Dornheim_PRL_2020,Moldabekov_JCTC_2022,Dornheim_review}. However, this procedure requires individual simulations for multiple perturbation amplitudes for each value of $q$, which rules out extensive parameter scans. A more convenient and more elegant alternative is given by the imaginary-time version of the fluctuation--dissipation theorem~\cite{Dornheim_MRE_2023}
\begin{eqnarray}\label{eq:static_chi}
    \chi(\mathbf{q}) = - n \int_0^\beta \textnormal{d}\tau\ F(\mathbf{q},\tau)\ ,
\end{eqnarray}
which implies that one can obtain the full $q$-dependence from a single simulation of the unperturbed system from the imaginary-time density--density correlation function $F(\mathbf{q},\tau)=\braket{\hat{n}(\mathbf{q},\tau)\hat{n}(-\mathbf{q},0)}$. We find the equivalent convergence behavior with respect to $P$ for $\chi(\mathbf{q})$ as for $S(\mathbf{q})$, except in the limit of large $q$; here, the $P=10$ data sets drastically deviates from the rest. This, however, is not primarily a factorization error of the density operator $\hat{\rho}=e^{-\beta\hat{H}}$, but rather a quadrature error in the evaluation of the integral of Eq.~(\ref{eq:static_chi}), since $F(\mathbf{q},\tau)$ can only be estimated on the given $P$ imaginary-time slices. In Fig.~\ref{fig:P_convergence}, we present a similar convergence analysis for the potential (top) and kinetic energy (bottom). In both cases, the present choice of $P=200$ primitive imaginary-time propagators is fully sufficient to even resolve small differences in the respective observables. We thus safely conclude that $P=200$ can faithfully resolve even small effects due to the considered pair potentials, Eq.~(\ref{eq:Ewald_potential}) and Eq.~(\ref{eq:Yakub_potential}).

\begin{figure}\centering
\includegraphics[width=0.45\textwidth]{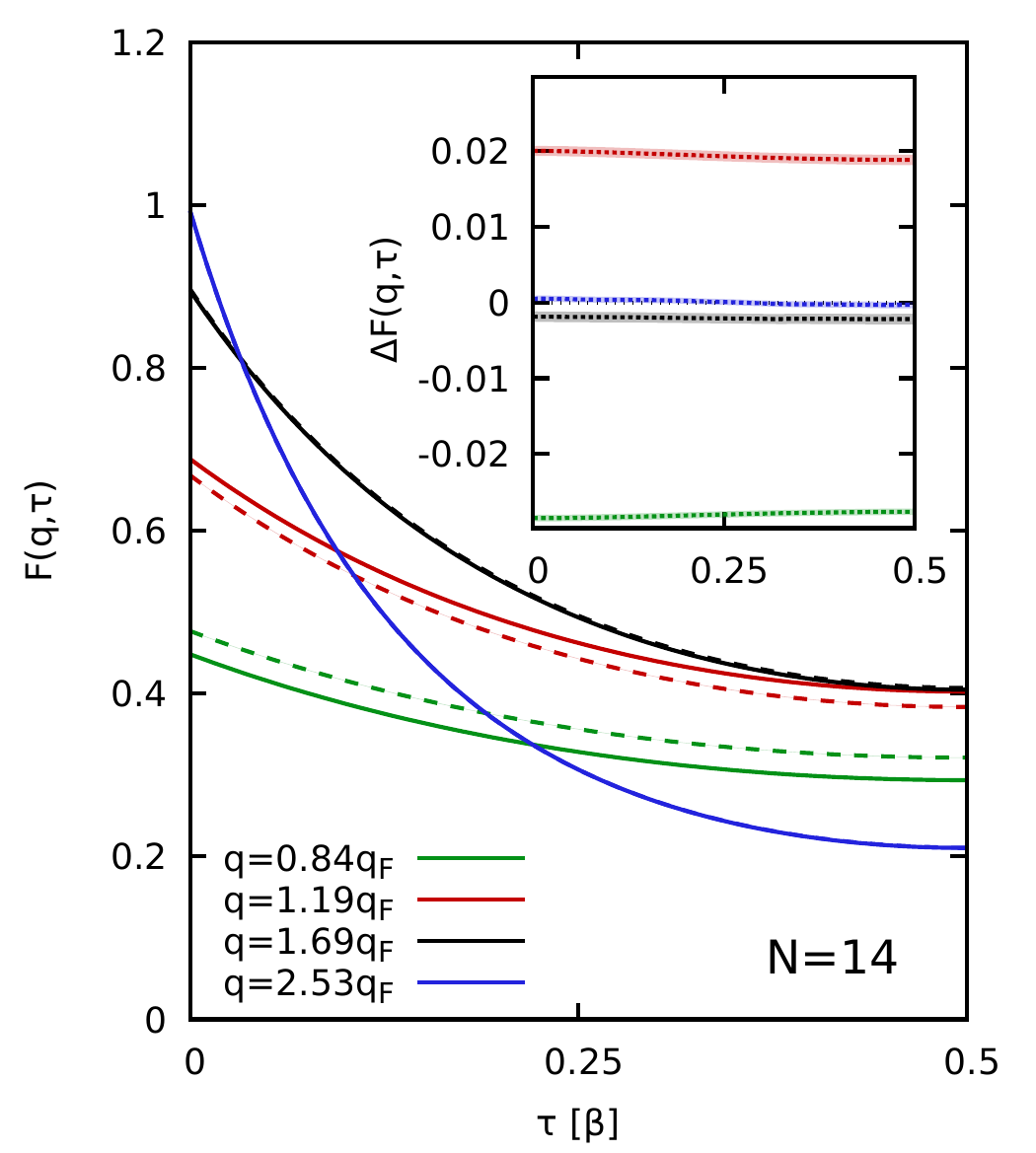}\\\vspace{-0.9cm}\includegraphics[width=0.45\textwidth]{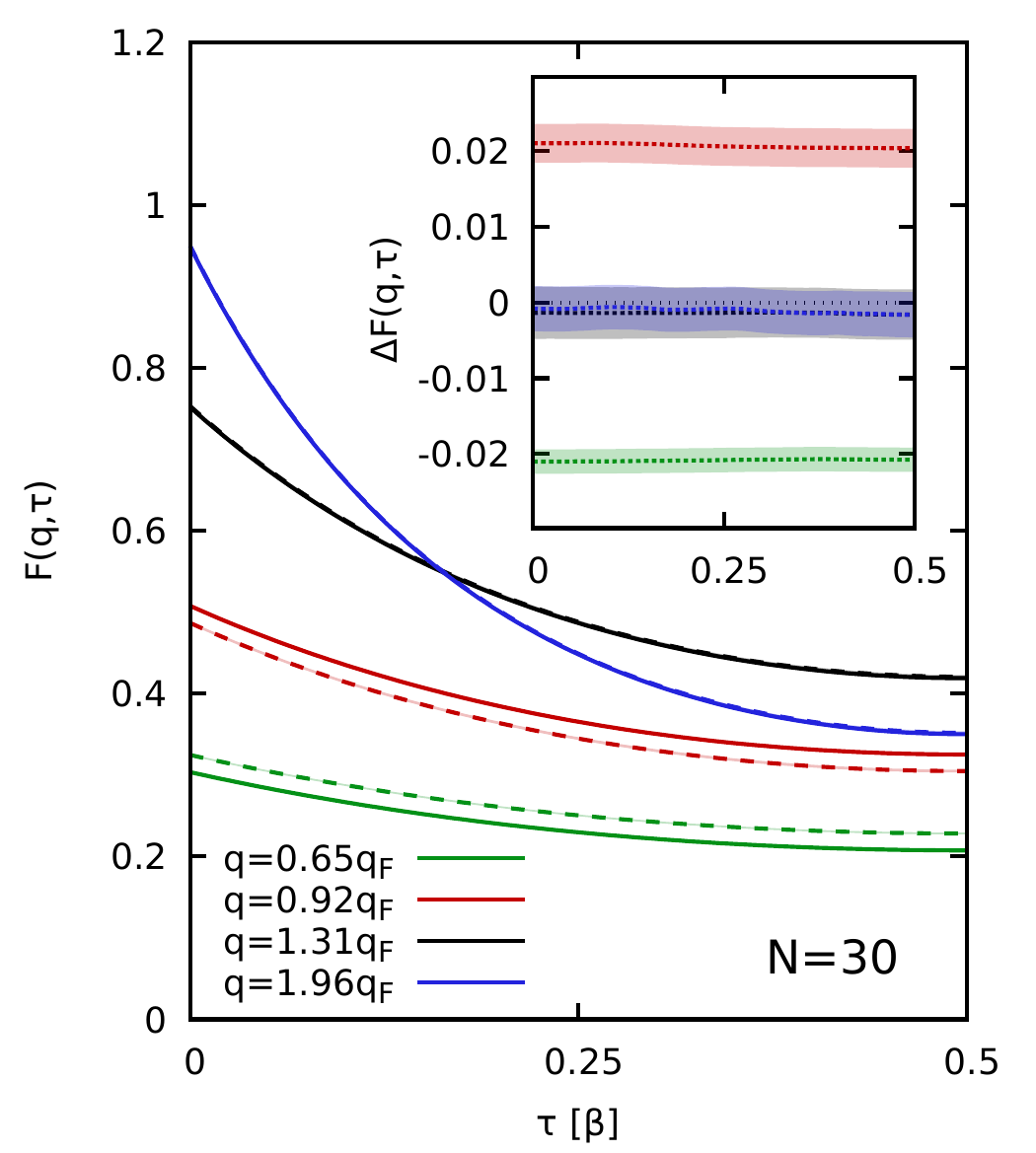}
\caption{\label{fig:ITCF_rs3p23_theta1} PIMC results for the density--density ITCF $F(\mathbf{q},\tau)$ of the UEG at $r_s=3.23$, $\Theta=1$ with $N=14$ (top) and $N=30$ (bottom). The solid and dashed lines have been computed using the Ewald and YR pair potentials, respectively. The insets show the absolute difference between the two, i.e., $\Delta F(\mathbf{q},\tau)=F_\textnormal{E}(\mathbf{q},\tau)-F_\textnormal{YR}(\mathbf{q},\tau)$. The shaded areas indicate the given statistical uncertainty. 
}
\end{figure} 
\begin{figure}\centering
\includegraphics[width=0.45\textwidth]{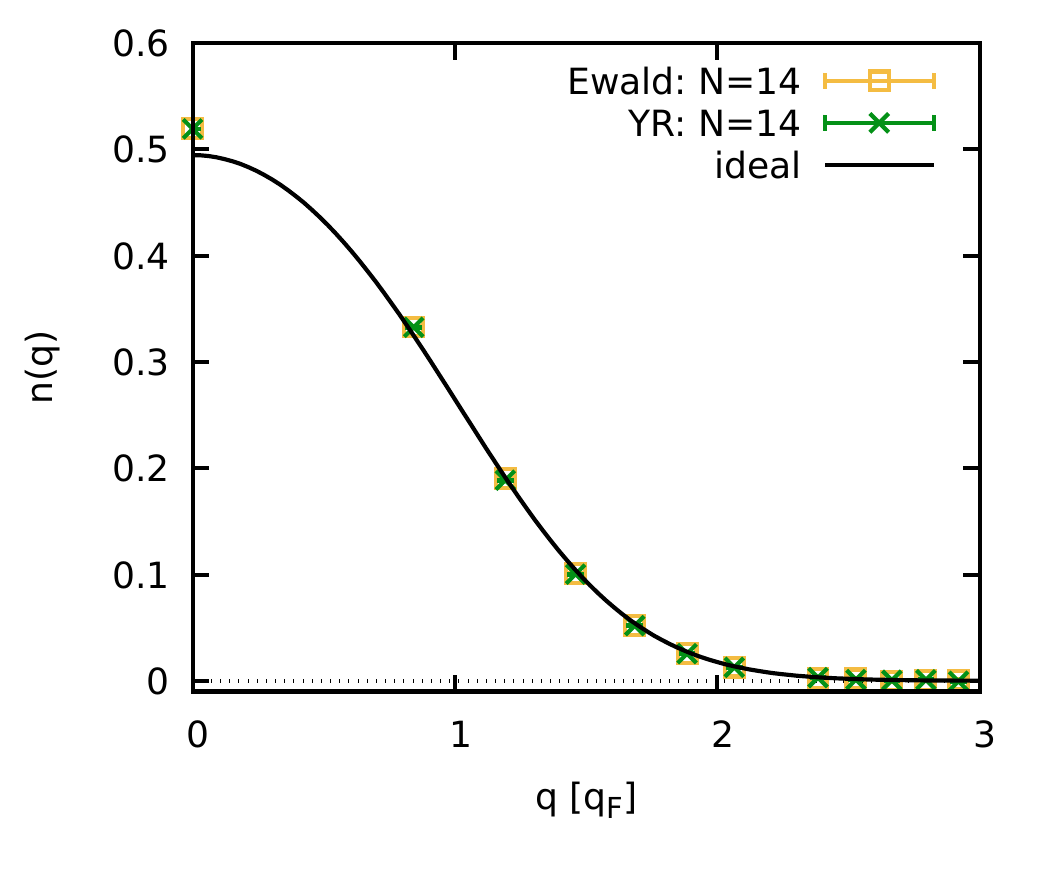}
\caption{\label{fig:Momentum_rs3p23_theta1} PIMC results for the momentum distribution $n(\mathbf{q})$ of the UEG with $N=14$ at $r_s=3.23$, $\Theta=1$. The yellow squares and green crosses have been computed using the Ewald and YR pair potentials, respectively. The solid black line corresponds to the Fermi distribution that describes the ideal Fermi gas at the same conditions.
}
\end{figure} 

Let us next come to the topic at hand, which is the comparison of simulation results between the Ewald potential [Eq.~(\ref{eq:Ewald_potential})] and the YR potential [Eq.~(\ref{eq:Yakub_potential})]. In the top row of Fig.~\ref{fig:SSF_rs3p23_theta1}, we compare the static structure factors, with the left and right panels corresponding to the former and the latter, respectively. The solid red curve has been computed within the \emph{effective static approximation} (ESA)~\cite{Dornheim_PRL_2020_ESA,Dornheim_PRB_ESA_2021}, which is known to perform well at these conditions, and has been included as a reference. The green crosses, black stars, and blue diamonds have been obtained from PIMC simulations of $N=14$, $N=20$, and $N=30$ unpolarized electrons, respectively. For completeness, we mention that we find an average sign of $S\approx0.05$ for $N=30$, precluding the simulation of much larger systems due to the fermion sign problem~\cite{dornheim_sign_problem}. There are no apparent finite size effects in the Ewald data (see also the magnified segment in the inset), except for the different $q$-grids; this is a well-known consequence of momentum quantization in the finite simulation cell~\cite{dornheim_prl}. This is expected based on previous investigations of $q$-resolved properties~\cite{Chiesa_PRL_2006,Drummond_PRB_2008,dornheim_prl,Holzmann_PRB_2016,Dornheim_JCP_2021}. In stark contrast, the results for $S(\mathbf{q})$ that have been obtained using the spherically averaged YR potential exhibit significant oscillations around the red ESA reference curve; this can again be discerned particularly well in the magnified inset.
The reason for these oscillations directly follows from the definition of the YR potential, Eq.~(\ref{eq:Yakub_potential}). Recall that particles around a distance of $r_\textnormal{cut}$ interact twice with a reference particle, cf.~Fig.~\ref{fig:scheme}. Consequently, the probability of finding two particles within such a distance gets reduced due to the additional potential energy penalty. The corresponding wavenumber $q_\textnormal{cut}=2\pi/r_\textnormal{cut}$ for $N=30$ electrons is indicated by the blue arrow at the bottom and by the vertical blue line in the inset of the top right panel of Fig.~\ref{fig:SSF_rs3p23_theta1}. As it is expected, the static structure factor is reduced in this segment (shaded blue area in the inset). This reduction is then compensated by the somewhat higher values of $S(\mathbf{q})$ around this segment, which explains the observed oscillations. Finally, we point out that the Ewald and YR potentials lead to indistinguishable results for $q\gg q_\textnormal{cut}$.

In the bottom row of Fig.~\ref{fig:SSF_rs3p23_theta1}, we repeat this analysis for the static linear density response function $\chi(\mathbf{q})$, cf.~Eq.~(\ref{eq:static_chi}). Overall, we find the same qualitative trends as for $S(\mathbf{q})$, i.e., no finite-size effects for Ewald and oscillations around the ESA reference curve for YR at $q\sim q_\textnormal{cut}$. Indeed, Eq.~(\ref{eq:static_chi}) implies a direct connection between the static structure factor and the linear static density response function as it holds $F(\mathbf{q},0)=S(\mathbf{q})$. The observed oscillations in $\chi(\mathbf{q})$ thus imply that the reduction in correlations around $q_\textnormal{cut}$ persist throughout the imaginary-time diffusion that is encoded in $F(\mathbf{q},\tau)$.

To further elucidate this observation, we show the full $\tau$-dependence of the ITCF in Fig.~\ref{fig:ITCF_rs3p23_theta1} for $N=14$ and $N=30$ electrons. In particular, the different colors correspond to different wavenumbers $q$, and the solid and dashed lines have been computed using the Ewald and YR potential, respectively; the associated Monte Carlo error bars are shown as shaded areas around these curves, but are barely visible due to the high data quality. We note that, in addition to its utility for the estimation of $\chi(\mathbf{q})$ via Eq.~(\ref{eq:static_chi}), the ITCF has attracted considerable recent attention. For example, it constitutes the starting point for the analytic continuation to compute the dynamic structure factor $S(\mathbf{q},\omega)$, see Sec.~\ref{sec:anal_cont} below, and contains rich information e.g.~about quasi-particle excitations~\cite{Dornheim_MRE_2023}, frequency moments of $S(\mathbf{q},\omega)$~\cite{Dornheim_moments_2023}, and quantum delocalization~\cite{Dornheim_PTR_2023}. Moreover, the ITCF has emerged as a standard tool for the interpretation of XRTS experiments and allows for the model-free estimation of the temperature~\cite{Dornheim_T_2022,Dornheim_T_follow_up,Schoerner_PRE_2023,Bellenbaum_APL_2025}, the absolute intensity~\cite{Dornheim_SciRep_2024}, and the Rayleigh weight~\cite{dornheim2024modelfreerayleighweightxray}.
Extensive discussions of its physical meaning have been presented in Refs.~\cite{Dornheim_PTR_2023,Dornheim_MRE_2023,Dornheim_review} and need not be repeated here.

Overall, we find that the effect of the YR potential onto $F(\mathbf{q},\tau)$ is nearly independent from $\tau$ at these conditions and manifests as a constant shift. This can be discerned particularly well in the insets, which illustrate the absolute difference $\Delta F(\mathbf{q},\tau)=F_\textnormal{E}(\mathbf{q},\tau)-F_\textnormal{YR}(\mathbf{q},\tau)$; they are constant within the given uncertainty intervals (shaded areas). The effect of these differences onto the analytic continuation is discussed in Sec.~\ref{sec:anal_cont} below.

\begin{figure}\centering
\includegraphics[width=0.45\textwidth]{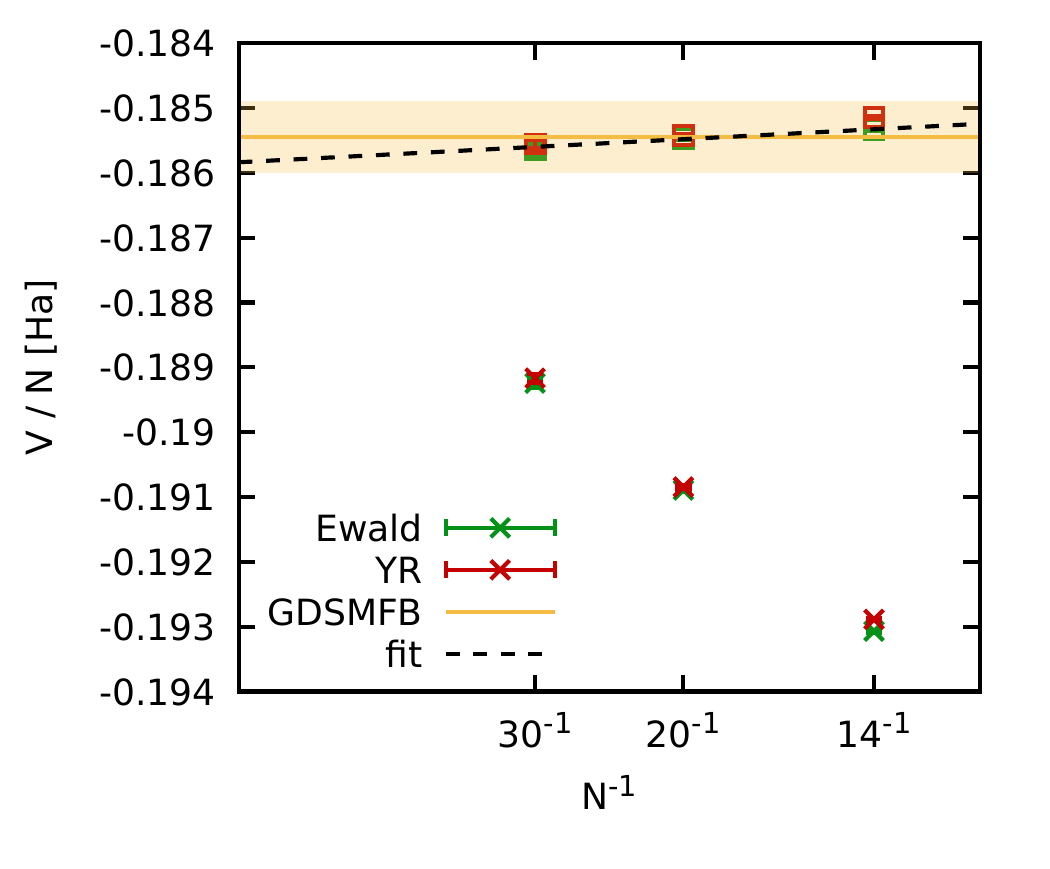}\\\includegraphics[width=0.45\textwidth]{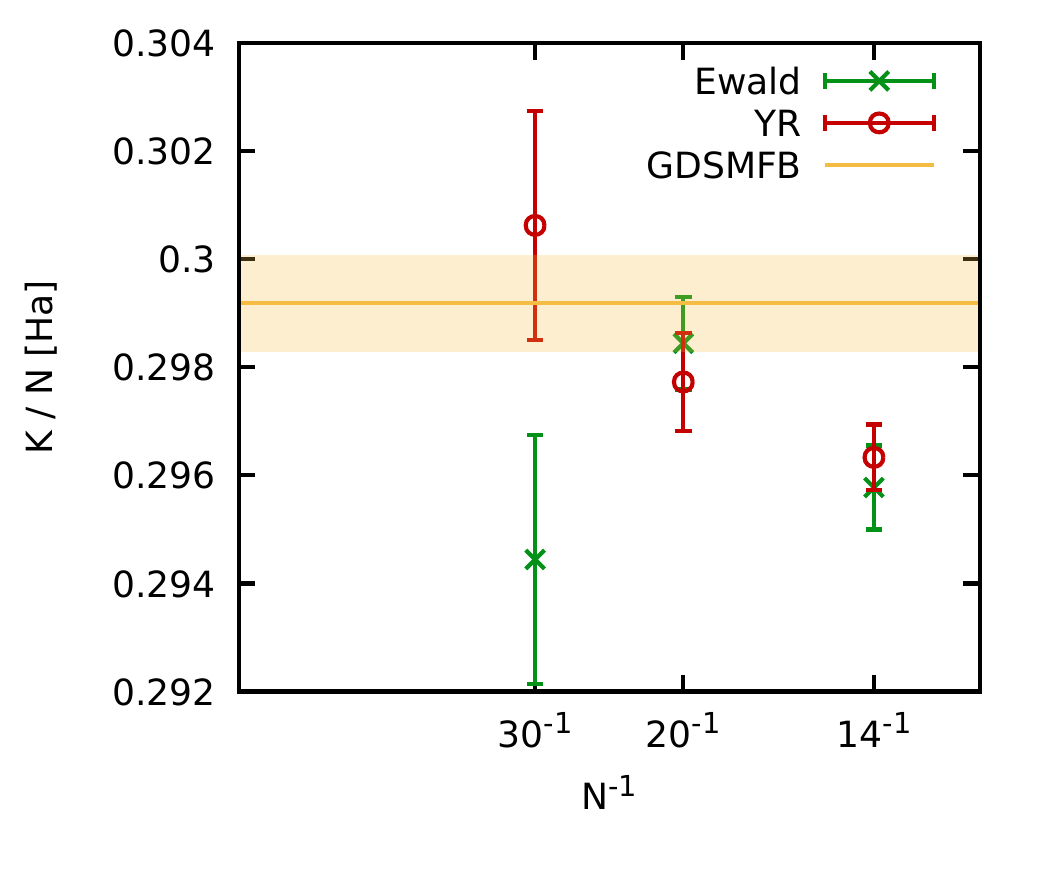}
\caption{\label{fig:Energy_rs3p23_theta1} (Top) Potential energy per electron $V/N$ computed with the Ewald (green) and YR pair potentials (red); crosses and squares show raw PIMC results, and finite-size corrected points via Eq.~(\ref{eq:BCDC}); the dashed black lines correspond to an empirical linear fit. (Bottom) Kinetic energy per electron $K/N$. All results are for the UEG at $r_s=3.23$, $\Theta=1$. The horizontal yellow lines and shaded yellow areas correspond to the parametrization by Groth \emph{et al.}~\cite{groth_prl} (GDSMFB) and its nominal uncertainty of $\pm0.3\%$.
}
\end{figure} 

\begin{figure*}\centering
\includegraphics[width=0.45\textwidth]{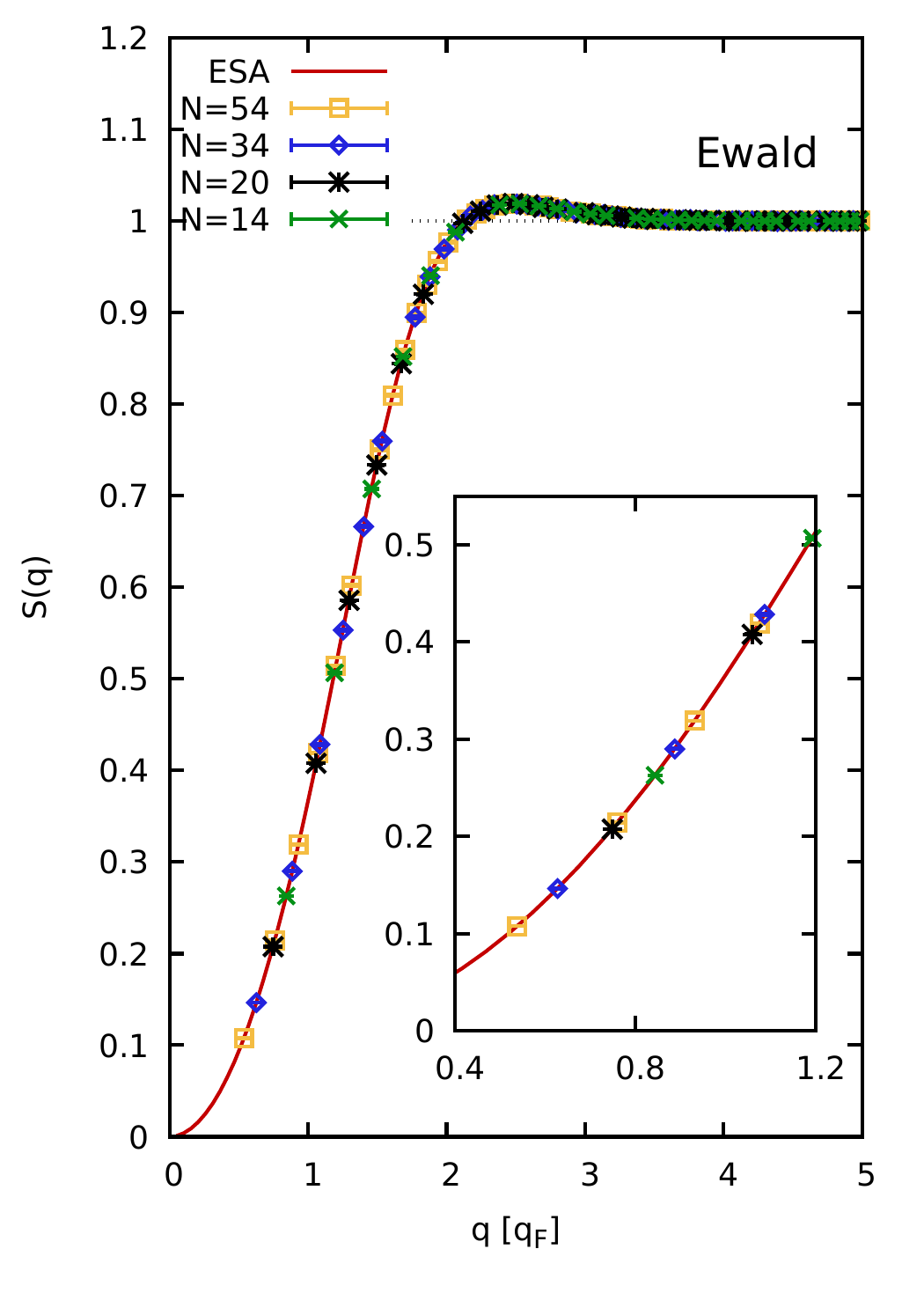}\includegraphics[width=0.45\textwidth]{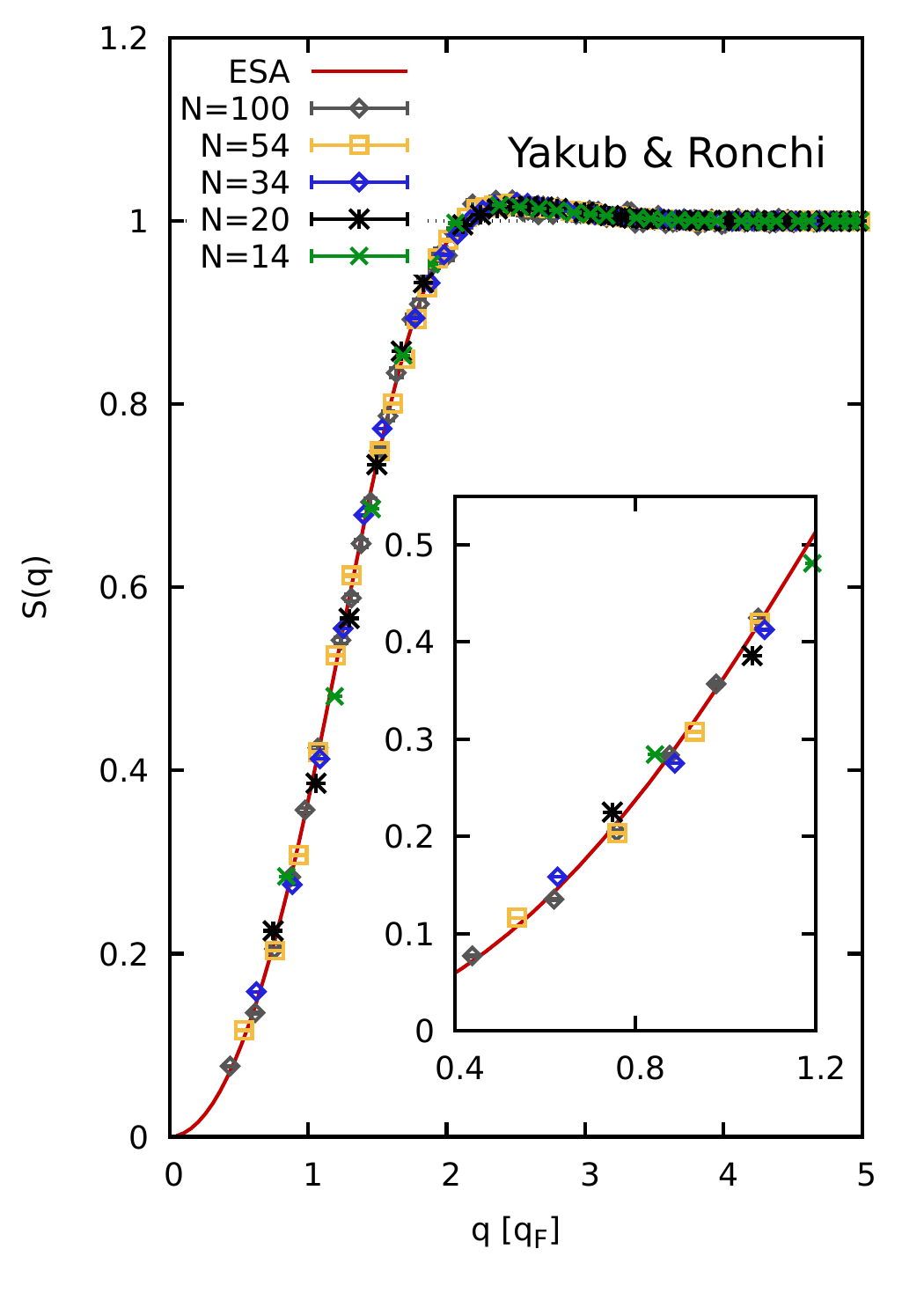}\\
\includegraphics[width=0.45\textwidth]{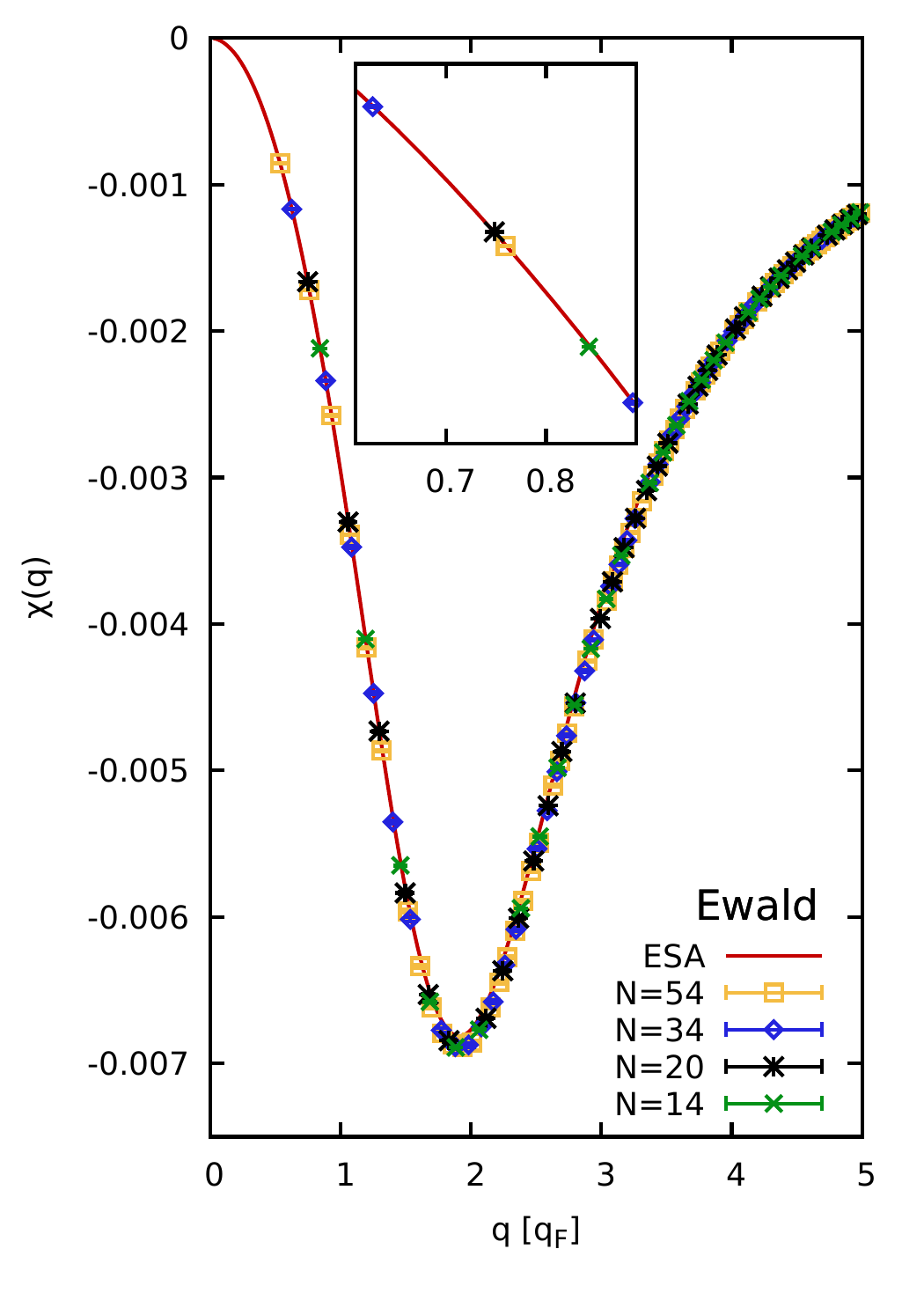}\includegraphics[width=0.45\textwidth]{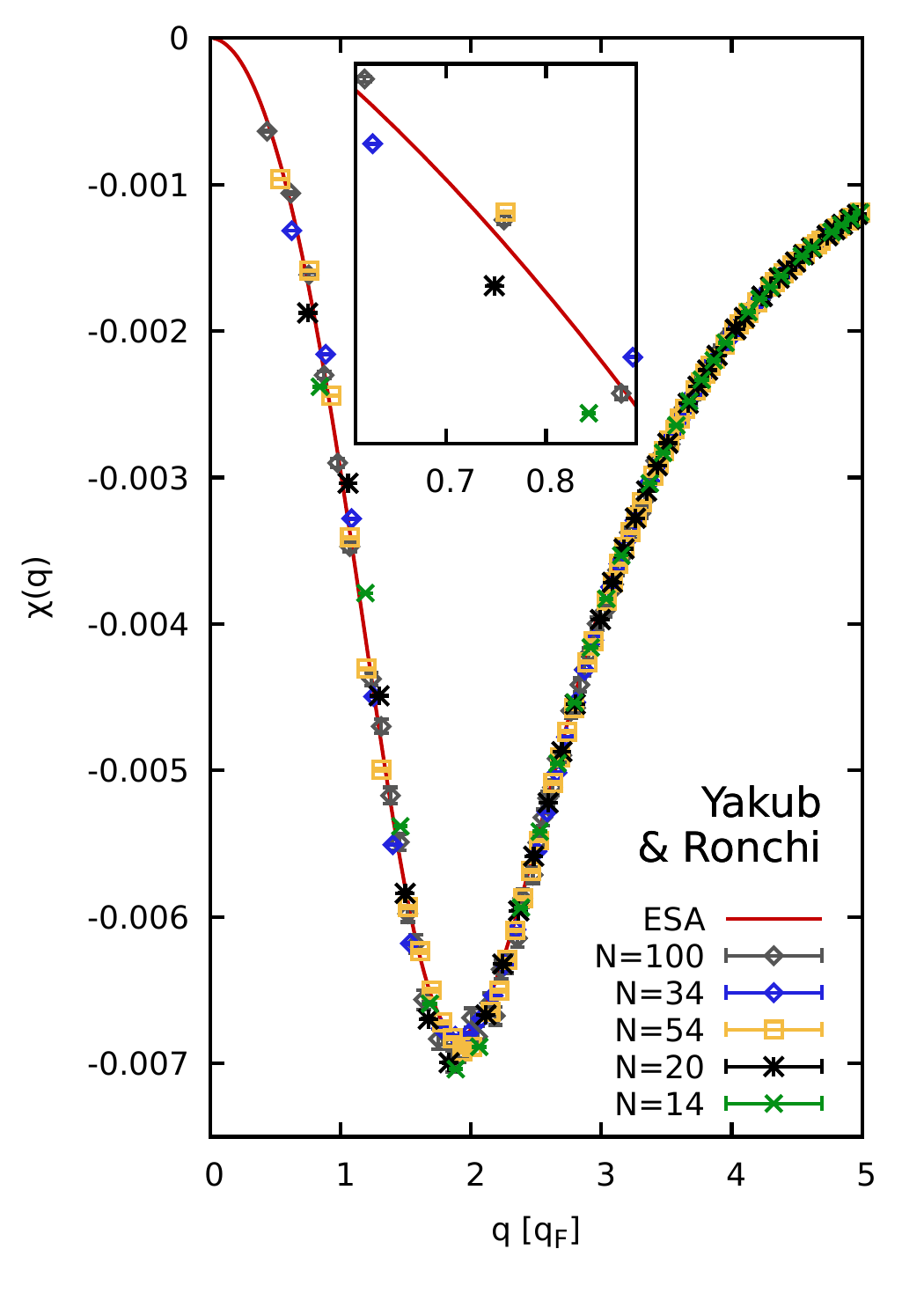}
\caption{\label{fig:SSF_rs10_theta1} PIMC results for the static structure factor $S(\mathbf{q})$ [top row] and the static linear density response function $\chi(\mathbf{q})$ [cf.~Eq.~(\ref{eq:static_chi}), bottom row] 
at $r_s=10$, $\Theta=1$. Symbols: PIMC data for different $N$ computed with the Ewald (left) and the YR potential (right). Solid red: effective static approximation (ESA)~\cite{Dornheim_PRL_2020_ESA,Dornheim_PRB_ESA_2021}, included as a reference. The insets show magnified segments.
}
\end{figure*} 

\begin{figure}\centering
\includegraphics[width=0.45\textwidth]{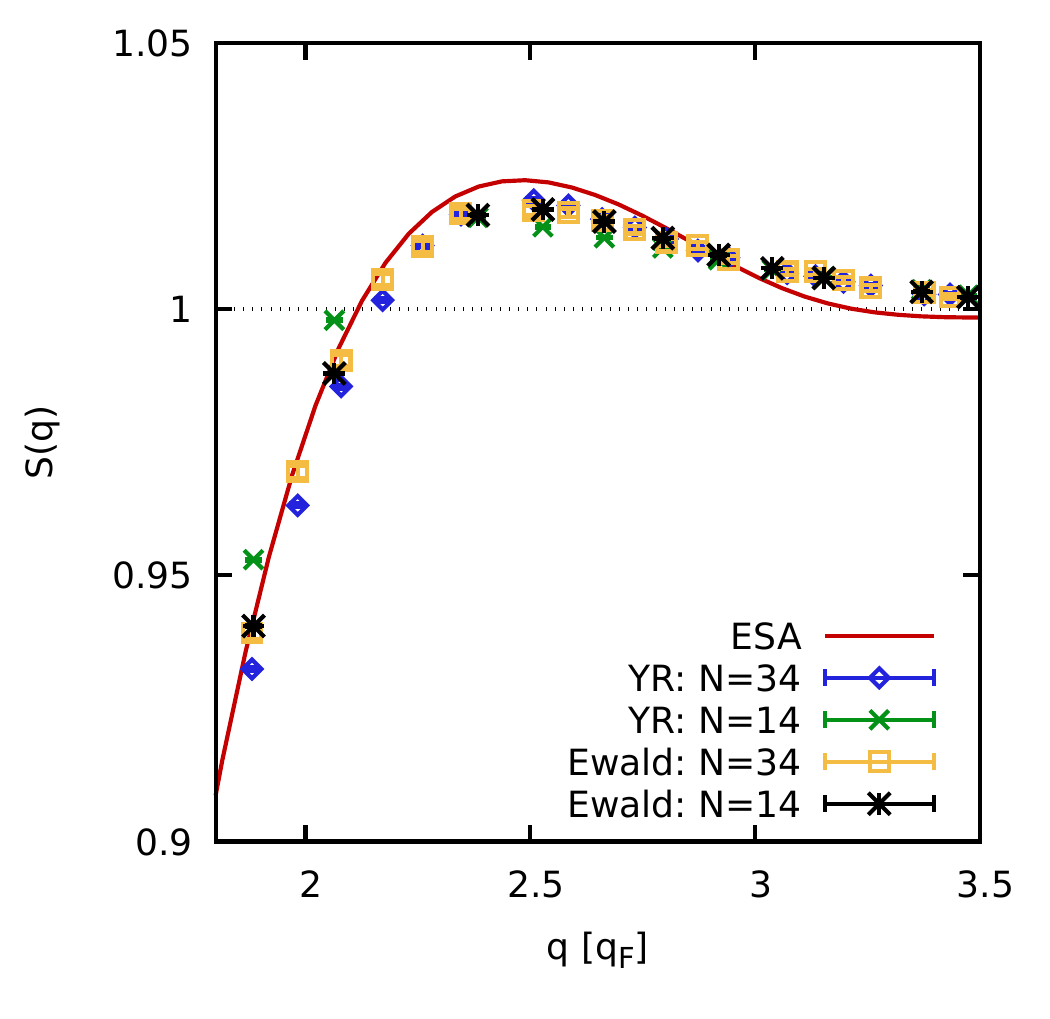}
\caption{\label{fig:zoom_rs10} Magnified segment from Fig.~\ref{fig:SSF_rs10_theta1} in the vicinity of the static structure factor maximum $S(\mathbf{q})$ at $r_s=10$ and $\Theta=1$. 
}
\end{figure} 

\begin{figure}\centering
\includegraphics[width=0.45\textwidth]{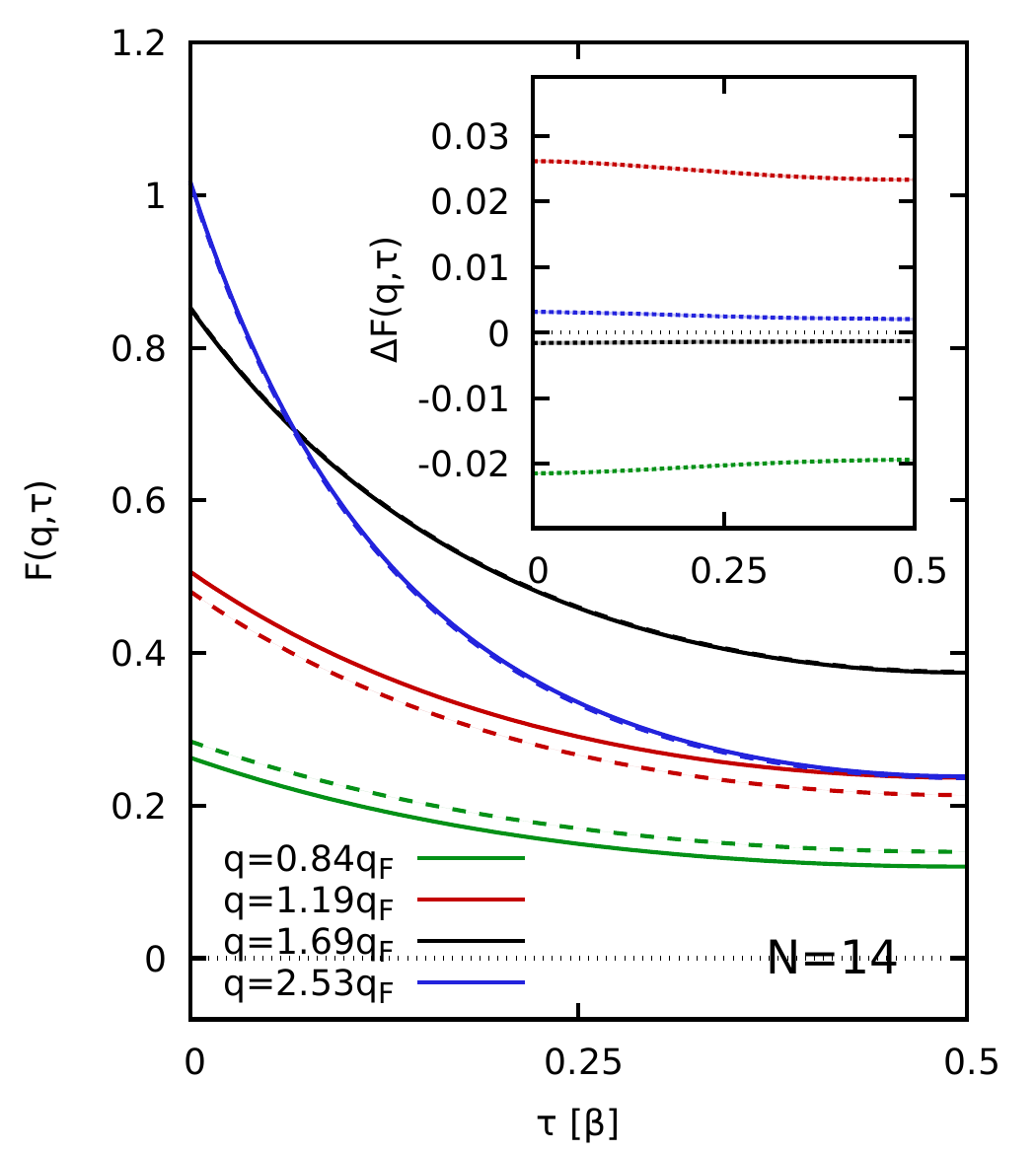}\\\vspace*{-0.9cm}\includegraphics[width=0.45\textwidth]{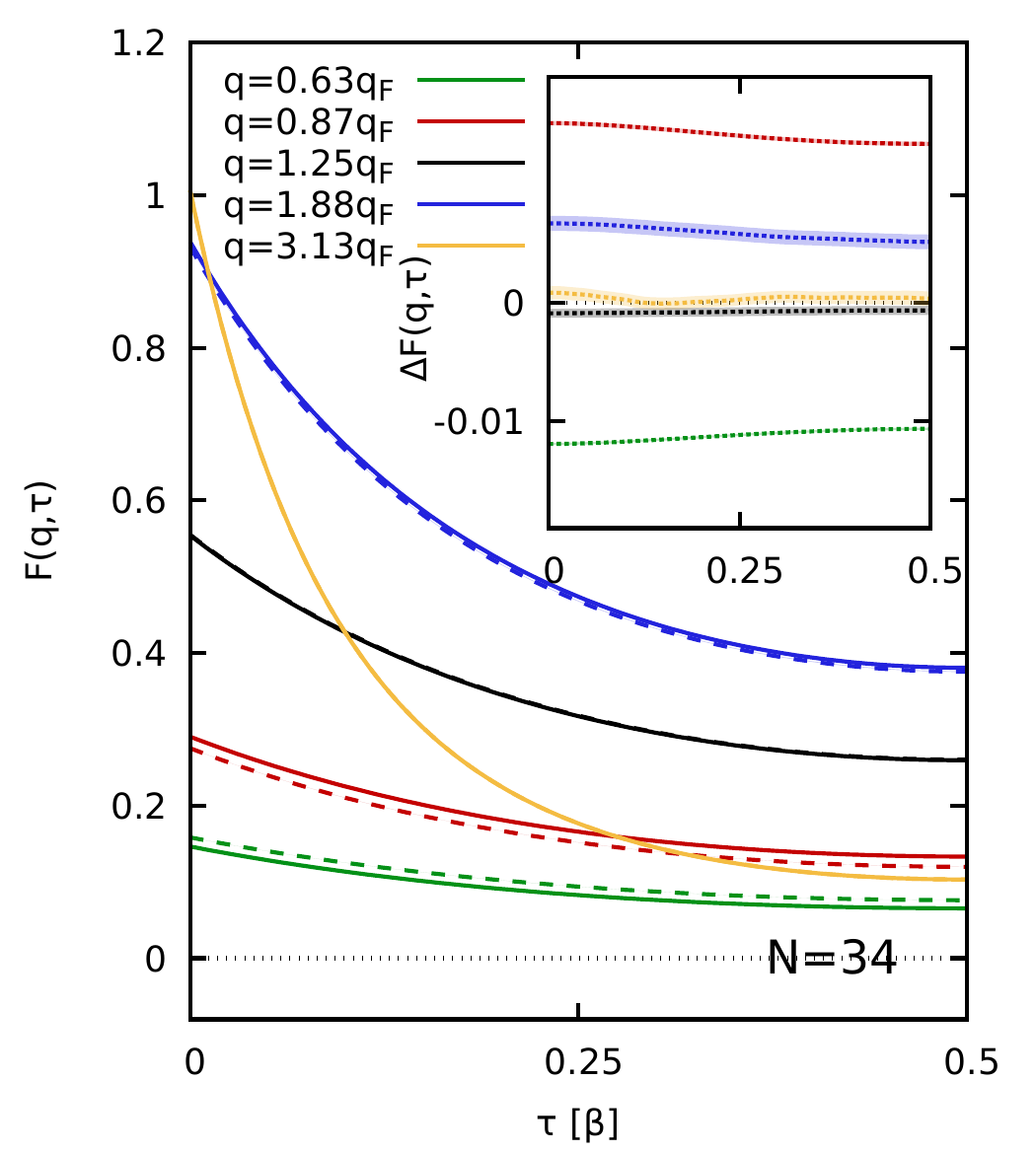}
\caption{\label{fig:ITCF_rs10_theta1} PIMC results for the density--density ITCF $F(\mathbf{q},\tau)$ of the UEG at $r_s=10$, $\Theta=1$ with $N=14$ (top) and $N=34$ (bottom). The solid and dashed lines have been computed using the Ewald and the YR pair potentials, respectively. The insets show the absolute difference between the two, i.e., $\Delta F(\mathbf{q},\tau)=F_\textnormal{E}(\mathbf{q},\tau)-F_\textnormal{YR}(\mathbf{q},\tau)$. The shaded areas indicate the given statistical uncertainty. 
}
\end{figure} 

An additional interesting question concerns the effect of the selected pair potential on the momentum distribution function $n(\mathbf{q})$. The latter is an off-diagonal observable within the PIMC formalism~\cite{cep,MILITZER201913} and can be conveniently estimated (including its proper normalization) using the extended ensemble approach introduced in Ref.~\cite{Dornheim_PRB_nk_2021}. The corresponding PIMC simulation results for $N=14$ are shown in Fig.~\ref{fig:Momentum_rs3p23_theta1}, with the yellow squares and green crosses obtained with the Ewald and YR pair potential, respectively. In addition, we have included the Fermi distribution function that describes the momentum distribution of the ideal Fermi gas (solid black curve) as a reference. Interestingly, we find an increase in the occupation of the zero-momentum state with respect to the ideal result; this is connected with a potential XC-induced lowering of the kinetic energy at finite temperatures for some densities, and has been discussed in detail in Refs.~\cite{Militzer_Pollock_PRL_2002,Hunger_PRE_2021,Dornheim_PRB_nk_2021,Dornheim_PRE_2021}. In the context of the present work, the key question concerns the effect of the pair potential; no differences between the Ewald and YR results can be resolved within the Monte Carlo error bars, even for $N=14$. In fact, this observation is not unexpected: $S(\mathbf{q})$ and $F(\mathbf{q},\tau)$ [from which we also estimate $\chi(\mathbf{q})$] are a two-body observables and, thus, subject to the double counting effect discussed above. In contrast, $n(\mathbf{q})$ is obtained from the Fourier transform of the single-particle density matrix, which is computed from correlations between the ends (usually denoted as \emph{head} and \emph{tail}) of an open imaginary-time trajectory. In particular, the \emph{head} and \emph{tail} do not interact with each other and are, thus not directly effected by the double counting.

Let us conclude the analysis of the warm dense UEG at $r_s=3.23$ and $\Theta=1$ by investigating the effect of the pair potential onto integrated properties. As two representative examples, we consider the potential and kinetic contributions to the energy in the top and bottom panels of Fig.~\ref{fig:Energy_rs3p23_theta1}, respectively. Both data sets (green and red crosses) exhibit the same qualitative behavior and we can only resolve small differences for $N=14$. This is unsurprising, as the potential energy is given by a sum over $S(\mathbf{q})$ and, hence, the oscillations in the latter cancel to a large degree. The green and red squares have been obtained by adding to the raw PIMC results the first-order finite-size correction that was first presented by Chiesa \emph{et al.}~\cite{Chiesa_PRL_2006} at $T=0$ and subsequently generalized by Brown \emph{et al.}~\cite{Brown_PRL_2013} to finite temperatures:
\begin{eqnarray}\label{eq:BCDC}
    \Delta V(N) = \frac{\omega_p}{4N}\textnormal{coth}\left(
\frac{\beta\omega_p}{2}
    \right)\ ;
\end{eqnarray}
the plasma frequency is given by $\omega_p=\sqrt{3/r_s^3}$. Evidently, Eq.~(\ref{eq:BCDC}) works well at the present conditions and removes most of the $N$-dependence from both data sets, see also the extensive discussion in Ref.~\cite{review}. We perform an empirical linear fit to the green squares, see the dashed black curve, which works well. Moreover, the extrapolation to the thermodynamic limit of $N^{-1}\to0$ lies within the $\pm0.3\%$ nominal uncertainty range (shaded yellow area) of the parametrization by Groth \emph{et al.}~\cite{groth_prl} (GDSMFB).

For the kinetic energy $K/N$ [bottom panel], no deviations between the two potentials can be resolved. This might be a consequence of either the reduced sensitivity of single-particle operators to two-body correlations (see Fig.~\ref{fig:Momentum_rs3p23_theta1} above), or simply be masked by the comparably larger error bars---a well known issue in PIMC simulations that originates from the large number of imaginary-time propagators $P$~\cite{Janke_JCP_1997}.
Overall, we observe that our PIMC results for $K/N$ are consistent with the GDSMFB result, although a conclusive quantitative statement is ruled out by the statistical uncertainty.

\subsection{Electron liquid\label{sec:electron_liquid}}

As a second example, we consider the UEG at $r_s=10$ and $\Theta=1$. These conditions are located at the boundary of the strongly coupled electron liquid regime~\cite{dornheim_electron_liquid,quantum_theory}, which gives rise to a roton-type feature in the dynamic structure factor $S(\mathbf{q},\omega)$ that has attracted significant recent attention~\cite{dornheim_dynamic,Dornheim_Nature_2022,Dornheim_Force_2022,koskelo2023shortrange,Takada_PRB_2016,Takada_PRB_2024}.

\begin{figure}\centering
\includegraphics[width=0.45\textwidth]{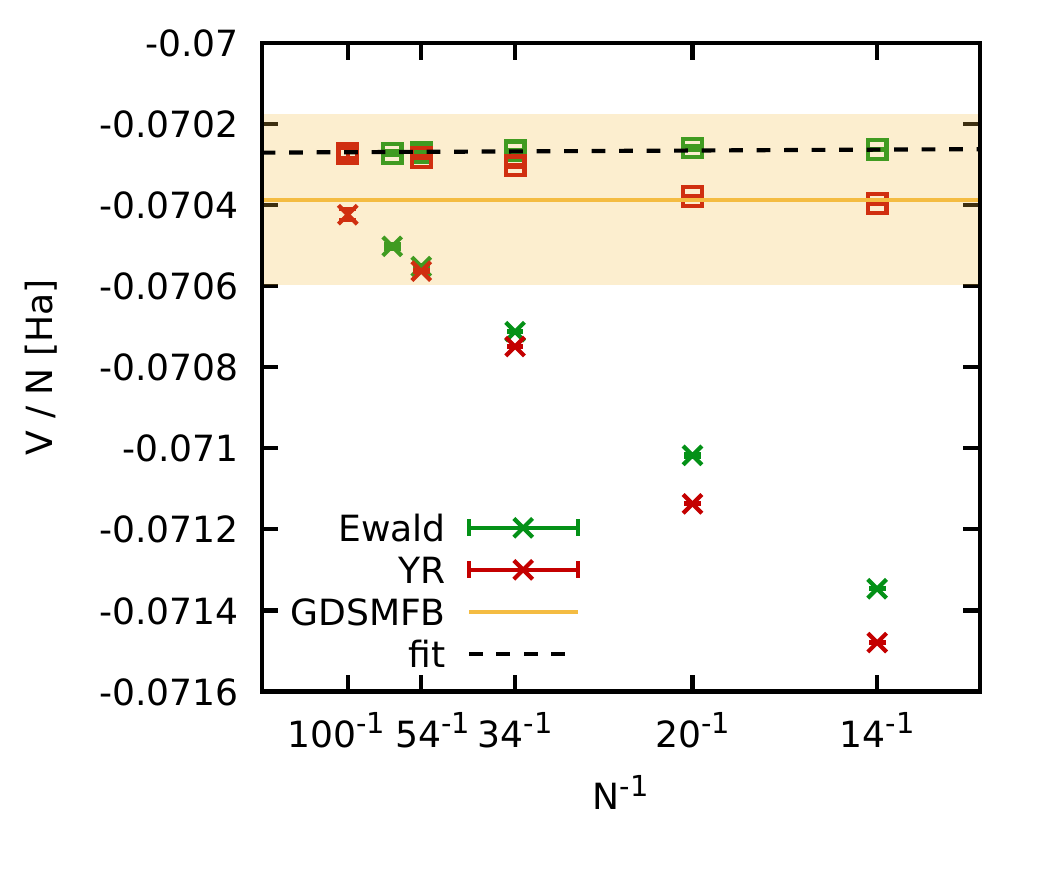}\\\includegraphics[width=0.45\textwidth]{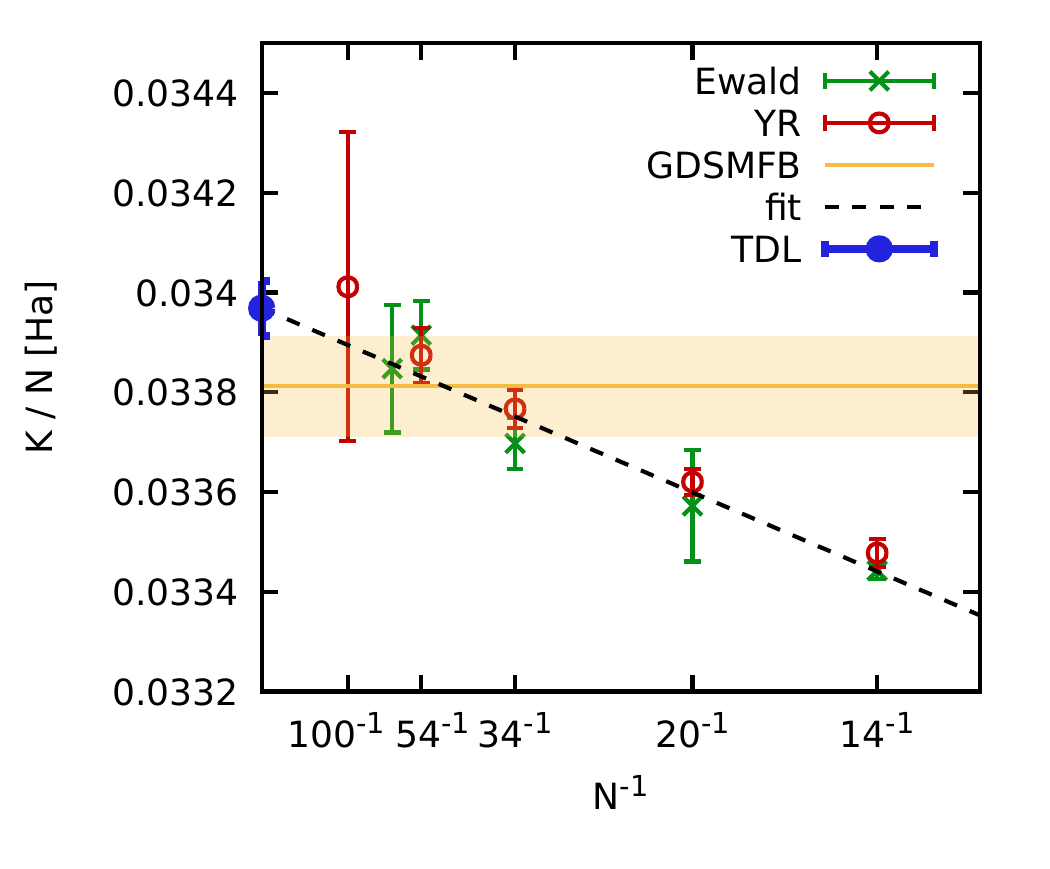}
\caption{\label{fig:Energy_rs10_theta1} (Top) Potential energy per electron $V/N$ computed with the Ewald (green) and the YR pair potential (red); crosses and squares show raw PIMC results, and finite-size corrected points via Eq.~(\ref{eq:BCDC}); the dashed black lines correspond to an empirical linear fit. (Bottom) Kinetic energy per electron $K/N$. All results are for the UEG at $r_s=10$, $\Theta=1$. The horizontal yellow lines and shaded yellow areas correspond to the parametrization by Groth \emph{et al.}~\cite{groth_prl} (GDSMFB) and its nominal uncertainty of $\pm0.3\%$.
}
\end{figure} 

In the top and bottom rows of Fig.~\ref{fig:SSF_rs10_theta1}, we analyze the effect of the potential onto $S(\mathbf{q})$ and $\chi(\mathbf{q})$, respectively, and find the same qualitative trends as for $r_s=3.23$.
In Fig.~\ref{fig:zoom_rs10}, we show a magnified segment around the maximum in $S(\mathbf{q})$. On this scale, we can clearly resolve small yet significant systematic errors in the ESA result. The YR potential is generally capable of resolving both the position and the height of the maximum, although small deviations from the Ewald results are visible even for $N=34$; the grey triangles showing the YR results for $N=54$, on the other hand, cannot be distinguished from the Ewald results for these $q$-values. In practice, the convergence of any YR data set with respect to $N$ thus has to be carefully
checked for any given $q$-range of interest.

Interestingly, the ITCF $F(\mathbf{q},\tau)$, which is analyzed in Fig.~\ref{fig:ITCF_rs10_theta1} is more strongly affected by the YR potential than at $r_s=3.23$ (cf.~Fig.~\ref{fig:ITCF_rs3p23_theta1} above).
In particular, the deviations $\Delta F(\mathbf{q},\tau)$ between the two results, which are shown in the insets, are no longer constant with respect to $\tau$, in particular for $N=14$. 
The implications of these findings for the analytic continuation to the dynamic structure factor $S(\mathbf{q},\omega)$ are investigated in Sec.~\ref{sec:anal_cont} below.

Finally, we analyze the potential and kinetic energies per particle in the top and bottom panels of Fig.~\ref{fig:Energy_rs10_theta1}. In the former, we find systematic deviations of $\sim0.1\%$ for $N\lesssim34$, which is still acceptable for many practical applications. In the latter, again, no systematic differences between the two pair potentials can be resolved. For completeness, we have performed an empirical linear fit of the Ewald result, see the dashed black line. The thus extrapolated result for the thermodynamic limit is shown by the bold blue point, and is consistent with GDSMFB within the given uncertainty.

\begin{figure*}
    \centering
    \includegraphics[width=0.45\linewidth]{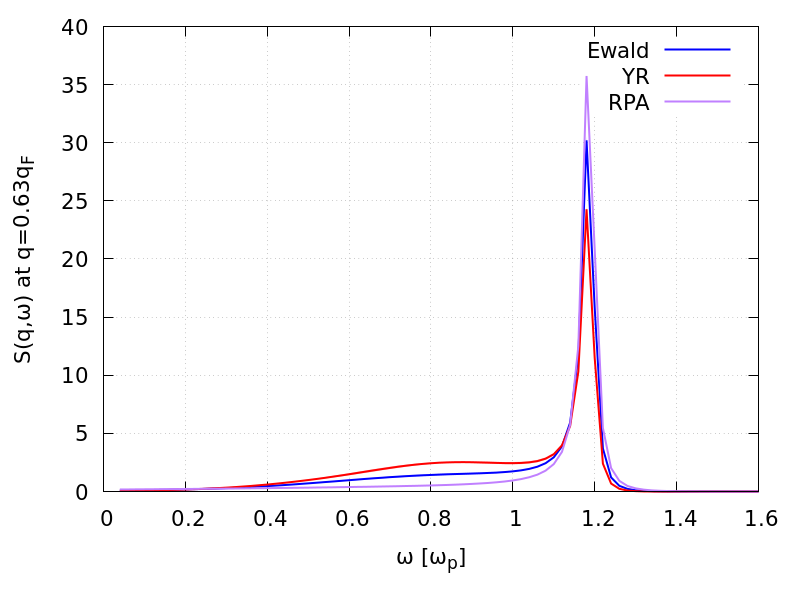}
    \includegraphics[width=0.45\linewidth]{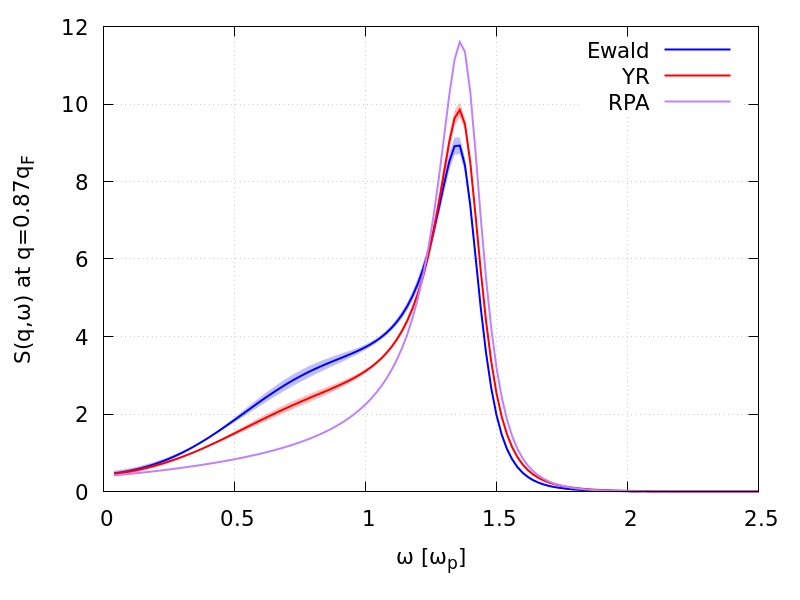}\\
    \includegraphics[width=0.45\linewidth]{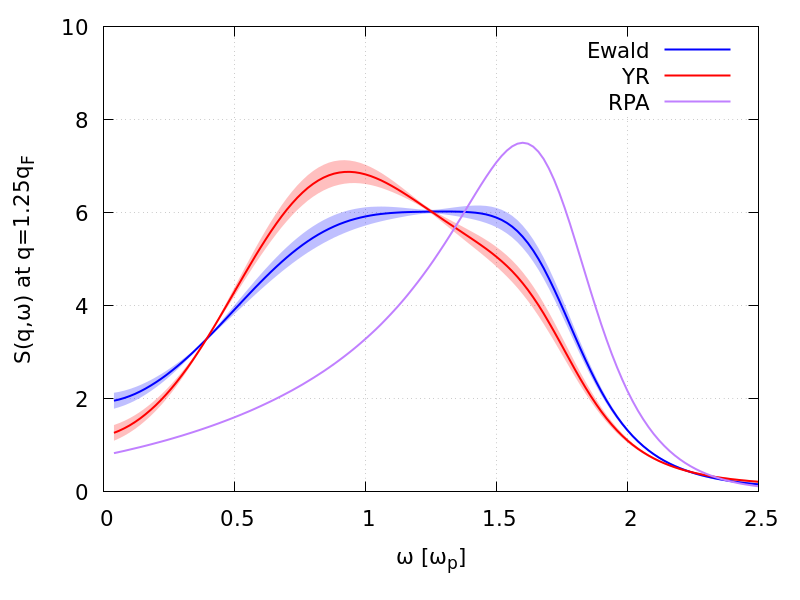}
    \includegraphics[width=0.45\linewidth]{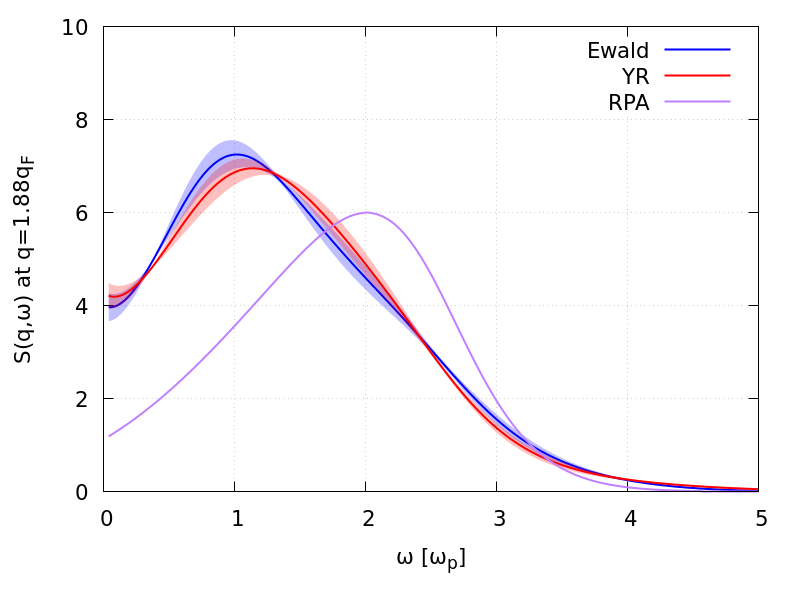}\\
    \includegraphics[width=0.45\linewidth]{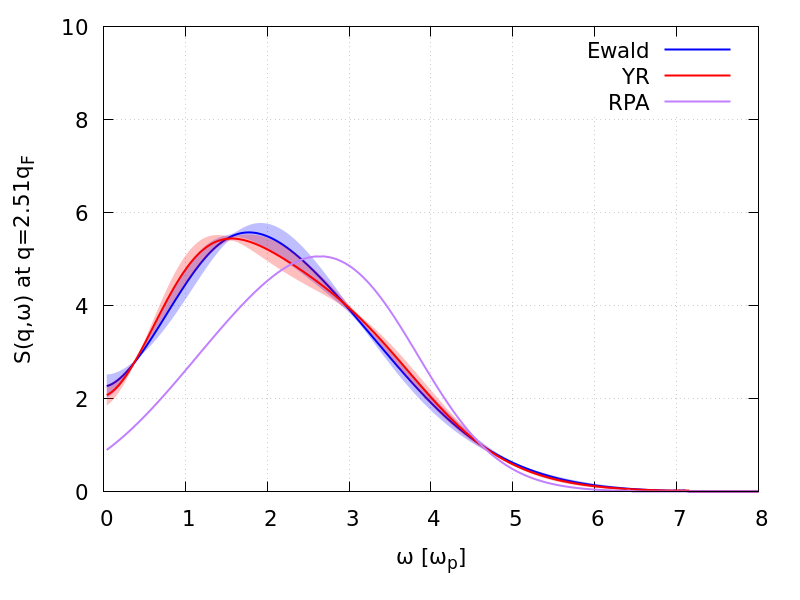}
    \includegraphics[width=0.45\linewidth]{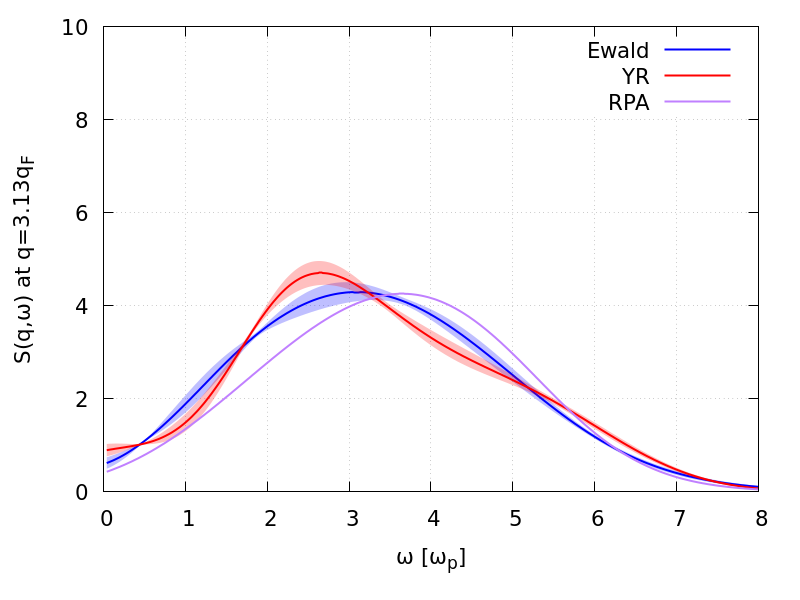}
    \caption{MEM estimates of $S(\mathbf{q},\omega)$, where $q$ values are indicated in the y-label. The MEM was conducted on ITCF data with either Ewald interaction (red) or YR interaction (blue). The RPA (purple) was used as the Bayesian prior in either case. At small $q$, statistically significant disagreement between the $S(\mathbf{q},\omega)$ obtained from Ewald and YR ITCF data.}
    \label{fig:MEM_Sqw}
\end{figure*}

\subsection{Sensitivity of analytic continuation\label{sec:anal_cont}}

In what follows, we analyze the effect of the pair potential onto analytic continuation results for the dynamic structure factor $S(\mathbf{q},\omega)$. The latter is connected to the ITCF via a two-sided Laplace transform~\cite{Dornheim_MRE_2023},
\begin{eqnarray}\label{eq:Laplace}
    F(\mathbf{q},\tau) = \int_{-\infty}^\infty \textnormal{d}\omega\ S(\mathbf{q},\omega)\ e^{-\tau\omega}\ ,
\end{eqnarray}
which needs to be inverted numerically; a notoriously difficult, exponentially ill-posed problem~\cite{JARRELL1996133}. For the UEG, Dornheim \emph{et al.}~\cite{dornheim_dynamic,dynamic_folgepaper,Dornheim_PRE_2020} have presented highly accurate results for $S(\mathbf{q},\omega)$ by incorporating a number of additional exact constraints~\cite{Dabrowski_PRB_1986}, many of which, however, are known only for the UEG but not for realistic WDM systems. Very recently, Chuna \emph{et al.}~\cite{chuna2025dualformulationmaximumentropy,chuna2025UEG}
have applied the all-purpose maximum entropy method (MEM) to the UEG and found that it gives very accurate results for $S(\mathbf{q},\omega)$ with high-quality PIMC input data for $F(\mathbf{q},\tau)$ and an appropriate choice for the Bayesian prior $\mu(\omega)$. Here, we consider $r_s=10$ and $\Theta=1$ for which we attain Monte Carlo error bars of an absolute magnitude of $\delta F(\mathbf{q},\tau)\sim10^{-5}$. Moreover, we choose the random phase approximation (RPA) for $\mu(\omega)$, which is a)  an appropriate choice for $r_s=10$~\cite{chuna2025UEG} and b) constitutes an equal prior for Ewald and YR potentials, allowing for a clear comparison between the two pair potentials.


\begin{figure}
    \centering
    \includegraphics[width=\linewidth]{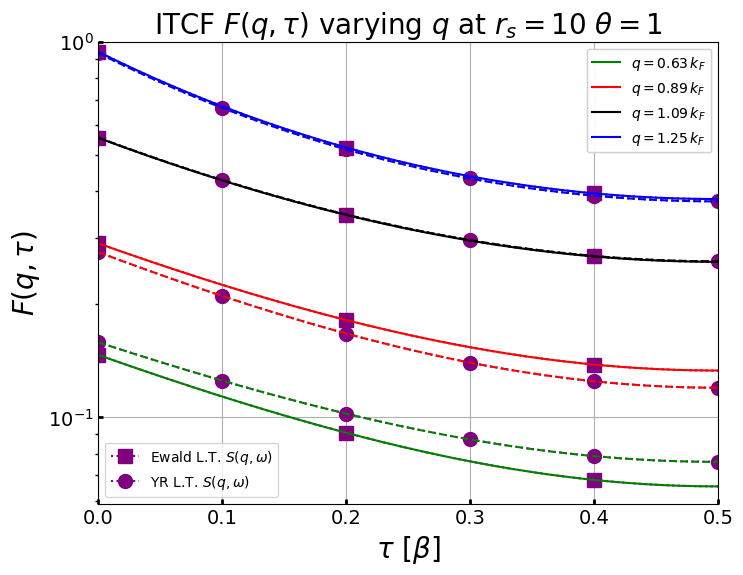}
    \caption{The ITCF data and the associated Laplace transformed MEM $S(\mathbf{q},\omega)$ estimate. The solid and dashed lines indicate ITCFs that have been computed using the Ewald and YR pair potentials, respectively. The MEM estimates associated to each ITCF (squares and circles) are indistinguishable from the respective data. For any MEM estimate, the $\chi^2$ value is of order $10^{-6}$ or less. 
    }
    \label{fig:MEM_ITCF_chi-square}
\end{figure}

In Fig.~\ref{fig:MEM_Sqw}, we show the thus reconstructed dynamic structure factors for six relevant wavenumbers $q$, with the blue and red curves corresponding to the Ewald and YR potentials. We note that the shaded areas around these curves show the uncertainty that is inherent in the inversion of Eq.~(\ref{eq:Laplace}), which takes into account both the statistical noise in the PIMC data for $F(\mathbf{q},\tau)$ via a leave-one-out method and the effect of averaging over regularization parameters in the entropy term that relates the reconstructed $S(\mathbf{q},\omega)$ to the default model $\mu(\omega)$; see Ref.~\cite{chuna2025UEG} for an extensive discussion of this procedure.
Finally, the purple curves show the RPA default model.

As a sanity check, we demonstrate that our MEM implementation perfectly reproduces the PIMC input data for four of the considered wavenumbers. This is illustrated in Fig.~\ref{fig:MEM_ITCF_chi-square}, where the solid and dashed lines correspond to the PIMC ITCF with Ewald and YR potentials, while the deep purple squares and crosses correspond to the ITCF as obtained by plugging the reconstructed dynamic structure factors into Eq.~(\ref{eq:Laplace}).

It is evident that there are significant differences in the MEM results for $S(\mathbf{q},\omega)$ itself. For $q=0.63q_\textnormal{F}$ (top left), we find that the utilization of the YR potential further dampens the plasmon, which is located at around $q\approx1.2\omega_{p}$ and which is already damped with the Ewald potential compared to RPA.
Moreover, YR leads to a substantial shift of spectral weight to lower frequencies. This is a direct consequence of the up-shift in the YR ITCF compared to the Ewald ITCF at this wavenumber, see Fig.~\ref{fig:ITCF_rs10_theta1}.
In fact, it is easy to see that a constant up-shift in the ITCF is associated with a contribution to the dynamic structure factor at $\omega=0$; this is highly relevant for two-component systems where the Rayleigh weight that describes the spectral weight of the ion feature manifests as such a shift~\cite{Dornheim_MRE_2024,bellenbaum2025estimatingionizationstatescontinuum}.
In the present case, the difference in the ITCF between the Ewald and YR calculations exhibits a weak dependence on $\tau$, which then leads to a shift of spectral densities to smaller, but still finite frequencies.

For $q=0.87q_\textnormal{F}$ (top right), we find the opposite trend. In this case, the YR plasmon is less damped than the Ewald result, and spectral weight is shifted to higher frequency. Again, this trend directly follows from the ITCF, in which the YR potential manifests as a down shift in this case.
In the middle row of Fig.~\ref{fig:MEM_Sqw}, we show results for $q=1.25q_\textnormal{F}$ (left) and $q=1.88q_\textnormal{F}$ (right), for which the impact of the pair potential is, however, less clear.
First, both wavenumbers are in the vicinity of the roton-type feature in the dispersion of the dynamic structure factor, and we observe a substantial exchange--correlation induced red-shift of spectral weight compared to the RPA default model for both Ewald and YR. 
For $q=1.25q_\textnormal{F}$, the PIMC results for the ITCF are very similar. Interestingly, we still observe quantitative differences in $S(\mathbf{q},\omega)$, which highlights the high sensitivity of the analytic continuation to any change in $F(\mathbf{q},\tau)$.
For $q=1.88q_\textnormal{F}$, $q=2.51q_\textnormal{F}$, and $q=3.13q_\textnormal{F}$, the effect of the pair potential is smaller and we can hardly resolve any significant differences within the given error bars.

\begin{figure}
    \centering
    \includegraphics[width=0.95\linewidth]{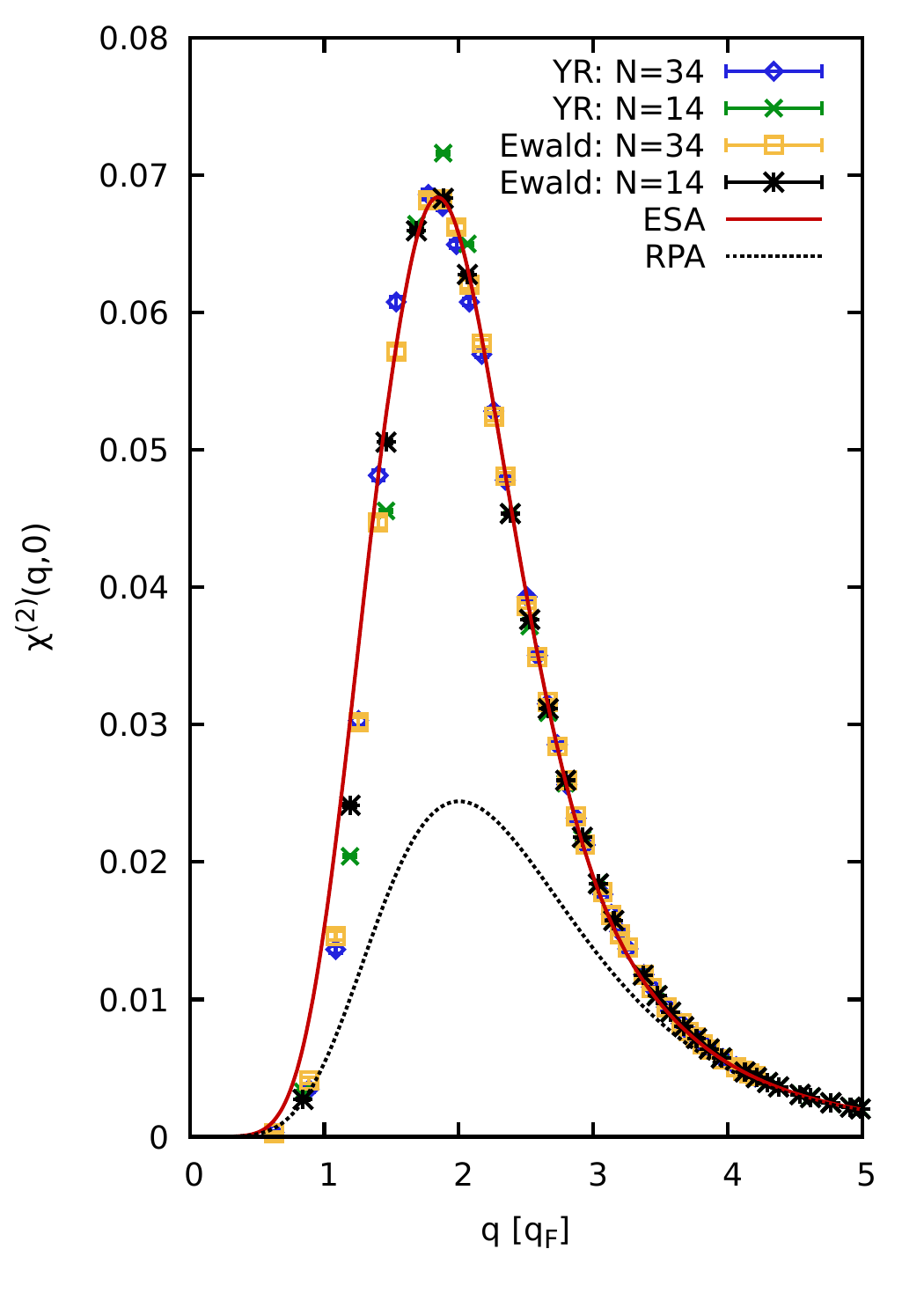}
    \caption{Static quadratic density response function [at the second harmonic of the original perturbation] $\chi^{(2)}(\mathbf{q},0)$ at $r_s=10$ and $\Theta=1$. Solid red (dotted black): analytical expressions from Ref.~\cite{Dornheim_PRR_2021} using as input the ESA local field correction~\cite{Dornheim_PRB_ESA_2021,Dornheim_PRL_2020_ESA} (within the random phase approximation). The symbols distinguish the Ewald and YR pair potentials, as well as $N=14$ and $N=34$ electrons.
    }
    \label{fig:quadratic}
\end{figure}

\subsection{Quadratic density response\label{sec:quadratic}}

While linear response theory is ubiquitous in WDM theory~\cite{Dornheim_review} and constitutes the basis for most diagnostics~\cite{siegfried_review,sheffield2010plasma,nolting}, there has emerged a recent interest in the nonlinear response properties of WDM~\cite{Dornheim_PRL_2020,Dornheim_PRR_2021,Dornheim_JCP_ITCF_2021,Dornheim_JPSJ_2021,Dornheim_CPP_2021,Dornheim_CPP_2022,Tolias_EPL_2023,vorberger2024greensfunctionperspectivenonlinear} (and beyond~\cite{mikhailov2012,Mikhailov_PRL}). Such effects are important e.g.~for stopping power calculations~\cite{PhysRevB.37.9268,PhysRevLett.122.015002,PhysRevB.56.15654,bergara1997,Kononov_PRB_2020} and for the construction of effective forces and potentials~\cite{Gravel,Dornheim_Force_2022}.
To our knowledge, the first reliable results for the nonlinear density response of the warm dense UEG have been presented in Ref.~\cite{Dornheim_PRL_2020} based on exact PIMC simulations of the harmonically perturbed system.
Subsequently, Dornheim \emph{et al.}~\cite{Dornheim_JCP_ITCF_2021} have generalized Eq.~(\ref{eq:static_chi}) to the non-linear density response by relating the latter to higher-order ITCFs.
For example, the static quadratic density response function $\chi^{(2)}(\mathbf{q},0)$ that describes the density response at the second harmonic of the original perturbation is given by
\begin{eqnarray}\label{eq:quadratic_response}
    \chi^{(2)}(\mathbf{q},0) = \frac{\Omega^2}{2} \int_0^\beta \textnormal{d}\tau_1 \int_0^\beta \textnormal{d}\tau_2\ F^{(2)}(\mathbf{q},\tau_1,\tau_2)\ ,
\end{eqnarray}
with the definition of the diagonal three-body ITCF
\begin{eqnarray}\label{eq:high}
    F^{(2)}(\mathbf{q},\tau_1,\tau_2) = \braket{ \hat{n}(2\mathbf{q},0) \hat{n}(-\mathbf{q},\tau_1) \hat{n}(-\mathbf{q},\tau_2) }\ .
\end{eqnarray}
In Fig.~\ref{fig:quadratic}, we analyze the effect of the pair potential onto PIMC results for $\chi^{(2)}(\mathbf{q},0)$.
The solid red and dotted black curves have been obtained using the analytical framework introduced in Ref.~\cite{Dornheim_PRR_2021}, which approximates screening effects on a linear level, using as input the ESA local field correction and random phase approximation (i.e., no local field correction), respectively.
Empirically, this framework works well for $q\gtrsim2q_\textnormal{F}$, but underestimates the true magnitude of screening effects for $q\to0$; in the context of the present work, its main purpose is as a guide to the eye.
The black stars and yellow triangles have been obtained using the Ewald pair potential and no dependence on the system size can be resolved within the given error bars. In contrast, the YR results (green crosses and blue diamonds) deviate from the Ewald results, and from each other for small to intermediate $q$.
First, we note that the particular dependence of these oscillations in $\chi^{(2)}(\mathbf{q},0)$ is less trivial compared to $S(\mathbf{q})$ and $\chi(\mathbf{q},0)$ investigated above, as it involves correlations between triples of particles instead of pairs, see Eq.~(\ref{eq:high}).
Second, we find significant inaccuracies in the vicinity of the peak in particular for $N=14$, which further substantiates our earlier recommendation for thorough $N$-comparisons in YR PIMC calculations.

\section{Summary and Outlook\label{sec:outlook}}

In this work, we have investigated in detail the effect of using the spherically averaged periodic potential by Yakub and Ronchi~\cite{Yakub_JCP_2003,Yakub2005} onto \emph{ab initio} PIMC simulations of the warm dense UEG. Overall, the YR potential reduces the computational effort by one to two orders of magnitude, which is particularly important for emerging capabilities to perform PIMC simulations with large numbers of electrons~\cite{Dornheim_JPCL_2024,dornheim2025fermionicfreeenergiestextitab}.

For integrated properties such as the kinetic and potential energy contributions, we find that the YR potential gives results in very close agreement with the more expensive Ewald potentials even at relatively strong coupling, $r_s=10$. This agrees with previous investigations~\cite{Levashov_CPP_2024,Demyanov_2022,Filinov_PRE_2023} and makes the YR potential the method of choice for such applications.

The situation is, however, more complex for $q$-resolved properties such as the static structure factor $S(\mathbf{q})$ and the static linear density response function $\chi(\mathbf{q})$. In particular, the YR potential effectively penalizes certain pair distances due to a double counting procedure, which introduces fluctuations in $q$-resolved properties in the vicinity of the cutoff wavenumber $q_\textnormal{cut}=2\pi/r_\textnormal{cut}$, where $r_\textnormal{cut}\approx0.6 L$ is of the same order as the length of the simulation cell. In practice, this means that the YR potential does not allow us to reliably study the small-$q$ limit, which is important for the description of XRTS experiments in a forward scattering geometry~\cite{siegfried_review,Bellenbaum_APL_2025,Sperling_PRL_2015,Gawne_PRB_2024}, the evaluation of compressibility sum rules~\cite{quantum_theory,Plagemann_PRE_2015,dharmawardana2025xraythomsonscatteringstudies}, and the estimation of optical and transport properties~\cite{GRABOWSKI2020100905}.

In addition, we have applied the recent maximum entropy method set-up by Chuna \emph{et al.}~\cite{chuna2025dualformulationmaximumentropy,chuna2025UEG} to PIMC results for the ITCF $F(\mathbf{q},\tau)$ with both Ewald and YR potentials. The thus obtained dynamic structure factors $S(\mathbf{q},\omega)$ exhibit significant deviations at small $q$, which further emphasizes the limitations of YR in this regime. On the other hand, the deviations vanish for $q\gtrsim2q_\textnormal{F}$, which might facilitate future YR calculations to describe and interpret XRTS measurements at large scattering angles that effectively probe the non-collective or even the single-particle regime~\cite{Dornheim_Science_2024,bellenbaum2025estimatingionizationstatescontinuum}.

Finally, we have analyzed the effect of the pair potential on the static quadratic density response at the second harmonic. This involves the PIMC estimation of an imaginary-time triple correlation function, leading to a less obvious effect of the YR potential. Most notably, we find a larger effect of the latter onto $\chi^{(2)}(\mathbf{q},0)$ in the vicinity of its maximum compared to the static linear response function $\chi(\mathbf{q})$.

Future efforts might include the application of the YR potential to PIMC simulations of real WDM systems that consist of both electrons and ions~\cite{Dornheim_MRE_2024,Dornheim_JCP_2024,Dornheim_Science_2024,Filinov_PRE_2023,Militzer_PRE_2021,Driver_PRE_2018,Driver_PRL_2012,Bohme_PRL_2022}, combined with a dedicated treatment of the diverging Coulomb attraction at small distances~\cite{MILITZER201688,Bohme_PRE_2023,DEMYANOV2024109326}. Moreover, the YR potential can be easily combined with other optimization schemes such as path contraction~\cite{PhysRevE.93.043305} or an improved Metropolis treatment of long-range potentials~\cite{Janke_PRX_2023}.

\begin{acknowledgements}

\noindent This work was partially supported by the Center for Advanced Systems Understanding (CASUS), financed by Germany’s Federal Ministry of Education and Research and the Saxon state government out of the State budget approved by the Saxon State Parliament. This work has received funding from the European Union's Just Transition Fund (JTF) within the project \emph{R\"ontgenlaser-Optimierung der Laserfusion} (ROLF), contract number 5086999001, co-financed by the Saxon state government out of the State budget approved by the Saxon State Parliament. This work has received funding from the European Research Council (ERC) under the European Union’s Horizon 2022 research and innovation programme (Grant agreement No. 101076233, "PREXTREME"). 
Views and opinions expressed are however those of the authors only and do not necessarily reflect those of the European Union or the European Research Council Executive Agency. Neither the European Union nor the granting authority can be held responsible for them. Computations were performed on a Bull Cluster at the Center for Information Services and High-Performance Computing (ZIH) at Technische Universit\"at Dresden and at the Norddeutscher Verbund f\"ur Hoch- und H\"ochstleistungsrechnen (HLRN) under grant mvp00024.
\end{acknowledgements}

\bibliography{bibliography}

\begin{thebibliography}{171}
\bibitem [{rep()}]{repo}%
  \BibitemOpen
  \href@noop {} {}\bibinfo {note} {A link to a repository containing all PIMC results will be made available upon publication.}\BibitemShut {Stop}%
\end{thebibliography}
\end{document}